\documentclass{jfm}

\usepackage{graphicx}
\usepackage{natbib}
\usepackage{multirow}
\usepackage{cite}
\usepackage{amsmath}
\usepackage{subfigure}

\newcommand{\ind}{\mbox{$\perp \kern-5.5pt \perp$}}

\newcommand{\eps}{\varepsilon}

\newcommand{\ud}{\mathrm{d}}

\newcommand{\abs}[1]{\left| #1 \right|}

\newcommand{\prd}{\hspace{0.04in}.}

\newcommand{\lp}{\left(}
\newcommand{\rp}{\right)}

\newcommand{\ba}{\begin{array}}
\newcommand{\bal}{\begin{array}{l}}
\newcommand{\ea}{\end{array}}

\newcommand{\om}{\omega}

\newcommand{\beq}{\begin{equation}}
\newcommand{\beqs}{\begin{equation*}}
\newcommand{\eeq}{\end{equation}}
\newcommand{\eeqs}{\end{equation*}}

\ifCUPmtlplainloaded \else
  \checkfont{eurm10}
  \iffontfound
    \IfFileExists{upmath.sty}
      {\typeout{^^JFound AMS Euler Roman fonts on the system,
                   using the 'upmath' package.^^J}%
       \usepackage{upmath}}
      {\typeout{^^JFound AMS Euler Roman fonts on the system, but you
                   dont seem to have the}%
       \typeout{'upmath' package installed. JFM.cls can take advantage
                 of these fonts,^^Jif you use 'upmath' package.^^J}%
      }
  \else
  \fi
\fi

\ifCUPmtlplainloaded \else
  \checkfont{msam10}
  \iffontfound
    \IfFileExists{amssymb.sty}
      {\typeout{^^JFound AMS Symbol fonts on the system, using the
                'amssymb' package.^^J}%
       \usepackage{amssymb}%
       \let\le=\leqslant  
       \let\ge=\geqslant  
      }{}
  \fi
\fi

\ifCUPmtlplainloaded \else
  \IfFileExists{amsbsy.sty}
    {\typeout{^^JFound the 'amsbsy' package on the system, using it.^^J}%
     \usepackage{amsbsy}}
    {\providecommand\mathbf[1]{\mbox{\boldmath $##1$}}}
\fi

\newsavebox{\astrutbox}
\sbox{\astrutbox}{\rule[-5pt]{0pt}{20pt}}

\title[Magma DSWs]{Dispersive Shock Waves in Viscously Deformable Media}

\author[N.K. Lowman and M.A. Hoefer]%
{N\ls I\ls C\ls H\ls O\ls L\ls A\ls S\ns K.\ns L\ls O\ls W\ls M\ls A\ls N%
  \thanks{Email address for correspondence: nklowman@ncsu.edu},\ns
\and M.\ns A.\ns H\ls O\ls E\ls F\ls E\ls R}

\affiliation{Department of Mathematics, North Carolina State University,
Raleigh, NC 27695, USA}

\pubyear{2010}
\volume{650}
\pagerange{119--126}
\date{}
\begin{document}

\maketitle


\begin{abstract}
  The viscously dominated, low Reynolds' number dynamics of
  multi-phase, compacting media can lead to nonlinear,
  dissipationless/dispersive behavior when viewed appropriately.  In
  these systems, nonlinear self-steepening competes with wave
  dispersion, giving rise to dispersive shock waves (DSWs).  Example
  systems considered here include magma migration through the mantle
  as well as the buoyant ascent of a low density fluid through a
  viscously deformable conduit.  These flows are modeled by a
  third-order, degenerate, dispersive, nonlinear wave equation for the
  porosity (magma volume fraction) or cross-sectional area,
  respectively.  Whitham averaging theory for step initial
  conditions is used to compute analytical, closed form predictions
  for the DSW speeds and the leading edge amplitude in terms of the
  constitutive parameters and initial jump height.  Novel physical
  behaviors are identified including backflow and DSW implosion for initial jumps sufficient to cause gradient
  catastrophe in the Whitham modulation equations.  Theoretical
  predictions are shown to be in excellent agreement with long-time
  numerical simulations for the case of small to moderate amplitude
  DSWs.  Verifiable criteria identifying the breakdown of this
  modulation theory in the large jump regime, applicable to a wide
  class of DSW problems, are presented.
\end{abstract}


\section{Introduction}
Shock waves in fluids typically arise as a balance between
nonlinearity and dissipatively dominated processes, mediated by the
second law of thermodynamics.  An alternative balancing mechanism
exists in approximately conservative media over time scales where
dissipation is negligible.  Nonlinearity and wave dispersion have been
observed to lead to dynamically expanding, oscillatory dispersive
shock waves (DSWs) in, for example, shallow water waves (known as
undular bores) \citep{chanson_tidal_2010}, ion-acoustic plasma (known as
collisionless shock waves)
\citep{taylor_observation_1970,tran_shocklike_1977}, superfluids
\citep{dutton_observation_2001,hoefer_dispersive_2006}, and ``optical
fluids'' \citep{wan_dispersive_2007,conti_observation_2009}.  It is
therefore counterintuitive that a fluid driven by viscous forces could
lead to shock waves regularized by dispersion.  In this work,
precisely this scenario is investigated in viscously deformable media
realized in magma transport and viscous fluid conduits.

The description of a low viscosity fluid flowing through a viscously
deformable, compacting medium is a fundamental problem in Earth
processes.  Such systems include flow of oil through compacting
sediment, subterranean percolation of groundwater through a fluidized
bed, or the transport of magma through the partially molten upper
mantle \citep{mckenzie84}.  This type of flow also has implications
for the buoyant ascent of a low density fluid through a deformable
vertical conduit.  Great interest by the broader scientific community
has been taken in these systems since the derivation of a set of
governing equations \citep{mckenzie84, scott84, scott86, fowler85}.  The
primitive equations treat the magma transport as flow of a low
Reynolds' number, incompressible fluid through a more viscous,
permeable matrix that is allowed to compact and distend, modeled as a
compressible fluid.  This model, thought to be a reasonable
representation of melt transport in the upper mantle, contrasts with
standard porous media flow where the matrix is assumed fixed and the
fluid is compressible.

Upon reduction to one-dimension (1D) and under a number of reasonable
simplifications, the model equations reduce to the dimensionless,
scalar magma equation \beq \label{eq: magma} \phi_t + (\phi^n)_z -
(\phi^n(\phi^{-m}\phi_t)_z)_z = 0, \eeq where $\phi$ is the porosity,
or volume fraction of the solid matrix occupied by the magma or melt,
and $(n,m)$ result from constitutive power laws relating the porosity
to the matrix permeability and viscosity, respectively.
\citet{scott84,scott86} concluded that the parameter space for
realistic magma systems is $n \in [2,5]$ and $m \in [0,1]$, a claim
which was later supported by re-derivation of the conservation
equations via homogenization theory for different geometric
configurations of the flow \citep{simpson10}.  The flow through a
deformable vertical conduit, magma migration via thermal plumes
through the convecting mantle being one example, can also be written
in the form \eqref{eq: magma} upon taking $(n,m) = (2,1)$ with the
interpretation of $\phi$ as the conduit's cross-sectional area
\citep{olson86}.

The magma equation \eqref{eq: magma} is a conservation law for the
porosity with nonlinear self-steepening due to buoyant advection
through the surrounding matrix via the flux term $\phi^n$ and nonlocal
dispersion due to compaction and distention of the matrix.  Solitary
traveling waves are special solutions to eq.~\eqref{eq: magma} that have
been studied in detail both theoretically \citep{scott84, scott86,
  richter84, barcilon86, nakayama92,simpson_asymptotic_2008} and
experimentally \citep{scott86b,olson86, helfrich90}.  A natural
generalization of single solitary waves to the case of a train of such
structures can be realized as a DSW when the porosity exhibits a
transition between two distinct  states.  The canonical
dynamical problem of this type is the determination of the long time
behavior of the dispersive Riemann problem, consisting of
eq.~\eqref{eq: magma} and the step initial data 
\beq \label{eq: ic_gen}
	\phi(z,0) =
	\begin{cases}
		\phi_-, & z \in (-\infty,0] \\
		\phi_+, & z \in (0,\infty)
      \end{cases} .  
\eeq
Note that there is currently no rigorous proof of well-posedness for
this particular initial value problem \citep{simpson07}.

\begin{figure}
  \centerline{\includegraphics{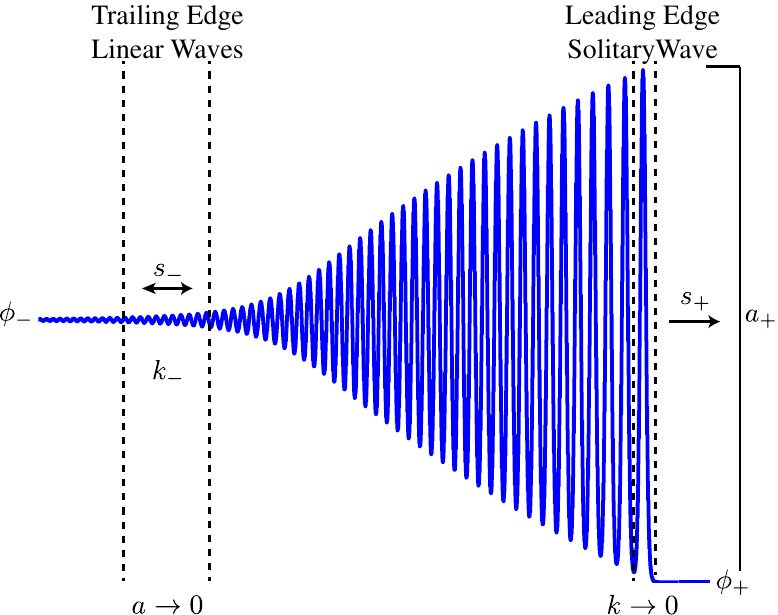}}
  \caption{Example numerical solution of the dispersive Riemann
    problem for eq. \eqref{eq: magma}.  The solution to the initial
    step connects the external constant states $\phi_\pm$ to an inner
    oscillatory region with a solitary wave leading edge where the
    wavenumber $k \rightarrow 0$ and a packet of linear waves in the
    trailing edge where $a \rightarrow 0$.  The salient physical
    properties of a magma DSW are the leading edge amplitude $a_+$,
    trailing edge wavenumber $k_-$, the forward propagation speed of
    the solitary wave front $s_+$, and the trailing edge speed $s_-$.}
\label{fig: schematic}
\end{figure}

The dispersive Riemann problem was first studied in numerical
simulations of eq. \eqref{eq: magma} in the $(n,m) = (3,0)$ case by
\citet{spiegelman93a,spiegelman93b}.  Rather than smoothing the
discontinuity and developing a classical shock front as in a
dissipatively regularized system, the magma system responds to a jump
by the generation of an expanding region of nonlinear oscillations
with a solitary wave front and small amplitude tail, characteristic of
DSWs (see figure \ref{fig: schematic}).  With the inclination to assume that steep gradients should be
regularized by dissipative processes in this viscous system,
\citet{spiegelman93b} used classical shock theory \citep{whitham74} to
attempt to describe the behavior.  In this work, we use a nonlinear
wave averaging technique \citep{el05}, referred to as Whitham
averaging \citep{whitham65}, in order to describe the dispersive
regularization of step initial data of arbitrary height with $0<
\phi_+ < \phi_-$ for a range of constitutive parameters $(n,m)$.  The
resulting DSW's leading and trailing edge speeds are determined and
the solitary wave front amplitude is resolved.  

Traditional analysis of DSWs, first studied in the context of the
Korteweg-de Vries (KdV) equation, asymptotically describes the
expanding oscillatory region via the slow modulation of a rapidly
oscillating, periodic traveling wave solution.  These modulations are
connected to the constant states of the exterior region by assuming
the presence of linear dispersive waves at one edge of the DSW
(amplitude $a \rightarrow 0$) and a solitary wave at the other
(wavenumber $k\rightarrow 0$), as shown in figure \ref{fig: schematic}
\citep{whitham65, gurevich74}.  In the case of KdV, the resulting
system of three hyperbolic modulation equations can be solved due to
the availability of Riemann invariants, which are, in general, not
available for non-integrable systems such as the one considered
here. An extension of simple wave led DSW Whitham modulation theory to
non-integrable systems has been developed by \citet{el05}, which has
been successfully applied, for example, to fully nonlinear, shallow
water undular bores \citep{el06, el09} and internal, two-fluid undular
bores \citep{esler11}.  The modulation equations reduce exactly to a
system of two hyperbolic equations at the leading and trailing edges
where Riemann invariants are always available.  Assuming the existence
of an integral curve connecting trailing and leading edge states in
the full system of modulation equations, one can calculate important
physical DSW properties, namely the edge speeds, the solitary wave
edge amplitude, and the trailing edge wavepacket wavenumber (see
$s_\pm$, $a_+$, and $k_-$ in figure \ref{fig: schematic}), with
knowledge of only the reduced system at the leading and trailing
edges.  We implement this Whitham-El simple wave DSW theory for the
magma dispersive Riemann problem, finding excellent agreement with
full numerical simulations in the small to moderate jump regime.  In
the large jump regime, we identify novel DSW behavior including
backflow (negative trailing edge speed, $s_-<0$) and DSW implosion.
The oscillatory region of the implosion is characterized by slowly
modulated periodic waves bookending an interior region of wave
interactions.  This behavior is associated with a change in sign of
dispersion and gradient catastrophe in the Whitham modulation
equations.  To the best of our knowlege, this is the first example of
breaking of the Whitham modulation equations for initial data of
single step type, previous studies having focused on breaking for
multistep initial data
\citep{grava_generation_2002,hoefer_interactions_2007,ablowitz_soliton_2009}
or an initial jump in a modulated periodic wave's parameters
\citep{jorge99}, resulting in quasi-periodic or multi-phase behavior.
Including gradient catastrophe, we identify four verifiable criteria
that can lead to the breakdown of the simple wave DSW theory,
applicable to DSW construction in other dispersive media.

Application of DSW theory to solutions of the magma equation has been
largely neglected in the previous literature.  \citet{elperin94}
considered the weakly nonlinear KdV reduction of the magma equation
and generic properties of the small amplitude DSWs produced.
\citet{marchant05} numerically integrated the full Whitham modulation
equations for the ``piston'' problem with $(n,m) =(3,0)$ incorporating
a Dirichlet boundary condition rather than considering the general Riemann
problem.  Furthermore, key DSW physical parameters were not discussed
in detail and neither backflow nor DSW implosion were
observed.  This work implements a general classification of weak to large amplitude DSW behavior in terms of the initial jump height and the constitutive parameters.

The presentation proceeds as follows.  Section \ref{sec: Governing Equations}
describes the derivation of the magma equation from the full set of
governing equations for both magma transport and viscous fluid
conduits.  Section \ref{sec: Equation Properties} presents properties
of the magma equation important to the application of Whitham theory.
Section \ref{sec: DSWs} implements Whitham theory to the
problem at hand and its comparison with numerical simulations is
undertaken in Section \ref{sec: discussion}.  Key physical
consequences of solution structures are elucidated and influences of
parameter variation are considered.  Causes of breakdown in the
analytical construction in the case of large jumps are identified.  We
conclude the manuscript with some discussion and future directions in
Section \ref{sec: summary}.

\section{Governing Equations}\label{sec: Governing Equations}
In this section, we briefly summarize the origin of the magma equation \eqref{eq: magma} in the context of magma migration \citep{mckenzie84, scott84, scott86} and fluid conduit flow \citep{olson86}.  Our purpose is to put the dispersive equation in its physical context in terms of assumptions, parameters, and scalings.


\subsection{Magma Geophysics} \label{sec: Geophysics} The equations
governing flow of a viscous interpenetrating fluid through a viscously
deformable medium were derived independently in the context of magma
by \citet{mckenzie84, scott84, scott86, fowler85}.  The system is a
generalization of standard rigid body porous media flow but exhibits
novel behavior.  In the absence of phase transitions, buoyancy drives
the predominant vertical advection of magma, the fluid melt, but the
inclusion of dilation and compaction of the solid matrix introduces
variability in the volume fraction occupied by the melt, which we will
see transmits melt fluxes through the system as dispersive porosity
waves.  In the interest of providing physical intuition for
understanding the 1D magma equation, we now recall the formulation of
McKenzie and describe the derivation of conservation equations for
mass and momentum of the fluid melt and the solid matrix, and then
reduce systematically to eq. \eqref{eq: magma}.  In what follows,
variable, primed quantities are unscaled, often dimensional.  Unprimed
variable quantities are all scaled and nondimensional.  Material
parameters and scales are unprimed as well.

The governing equations are a coupled set of conservation laws for
mass and momentum which describe the melt as an inviscid,
incompressible fluid and the solid matrix as a viscously deformable
fluid, written in terms of the porosity $\phi'$ (or volume fraction of
melt).  Interphase mass transfers are taken to be negligible so the
coupling comes from McKenzie's introduction of the interphase force
$\mathbf{I}'$, a generalization of the standard D'Arcy's law
\citep{scheidegger74}, which describes the rate of separation of the
melt and matrix as proportional to the gradients of the lithostatic
and fluid pressures.  The leading proportionality term is chosen so
that in the limit of a rigid matrix, D'Arcy's Law is recovered.  Upon
substitution of the coupling term into the governing equations, the
system reduces to the Stokes' flow equations for the matrix in the
``dry'' limit, $\phi' \rightarrow 0$.

To simplify clearly to the 1D magma equation, it is convenient to
write the governing equations as in \citet{katz07}, where the original
McKenzie system is presented in a computationally amenable form.  This
follows from taking the solid and fluid densities to be distinct
constants and then introducing a decomposition of the melt pressure
$P' = P_\mathrm{l}' + \mathcal{P}' + \left.P^*\right.'$ where
$P_\mathrm{l}' = \rho_\mathrm{s} g z'$ is the background lithostatic
pressure, $\mathcal{P}'$ is the pressure due to dilation and
compaction of the matrix given by $\mathcal{P}'=(\zeta' -
\frac{2}{3}\eta') \nabla \cdot \boldsymbol{v}'_{\mathrm{s}}$, and $
\left.P^*\right.'$ encompasses the remaining pressure contributions
primarily stemming from viscous shear stresses in the matrix.  Note
the introduction of the solid matrix velocity
$\boldsymbol{v}'_{\mathrm{s}}$, as well as the matrix shear and bulk
viscosities $\eta', \zeta'$, which arise due to matrix compressibility
and depend on the porosity as described below.  For
nondimensionalization, we follow the scalings described in
\citet{spiegelman93a} (a similar reduction was performed in
\citet{scott84, scott86, barcilon86, barcilon89}).  This requires the
introduction of the natural length scale of matrix compaction $\delta$
and the natural velocity scale of melt percolation $w_0$ proposed by
\citet{mckenzie84}, which for a background porosity $\phi_0$ are
defined as \beq \label{eq: scales} \delta = \sqrt{\frac{K'_0(\zeta'_0
    + \frac{4}{3}\eta'_0)}{\mu'}}, \quad w_0 = K'_0\frac{\Delta \rho
  g}{\phi_0 \mu} \eeq where $\mu$ is the melt viscosity, $\Delta \rho
= \rho_\mathrm{s} - \rho_\mathrm{f}$ is the difference between the
solid and fluid densities, and $K'_0$ and $(\zeta'_0 +
\frac{4}{3}\eta'_0)$ are the permeability and combined matrix
viscosity at the background porosity $\phi_0$, respectively.
\citet{spiegelman01} remarks that for practical geological problems,
$\delta$ is on the order of $10^2-10^4$ m while $w_0$ takes on values
of 1 - 100 $\mathrm{m~yr}^{-1}$, and the background porosity $\phi_0$
of standard media is between $10^{-3}-10^{-2}$.  Using these as
standard scales and after algebraic manipulation, the McKenzie system
reduces to the non-dimensional form (unprimed variables) of the system
presented in \citet{katz07}, \beq \label{eq: evolS} \frac{D\phi}{Dt} =
(1-\phi_0\phi)\frac{\mathcal{P}}{\xi}, \eeq \beq \label{eq: press}
-\nabla \cdot K \ \nabla \mathcal{P} + \frac{\mathcal{P}}{\xi} =
\nabla \cdot K \ [\nabla P^* + \mathbf{\hat{g}}] , \eeq
\beq \label{eq: cmpn} \nabla \cdot \boldsymbol{v}_{\mathrm{s}} =
\phi_0 \frac{\mathcal{P}}{\xi}, \eeq \beq \label{eq: shear} \nabla P^*
= \nabla \cdot \eta(\nabla \boldsymbol{v}_{\mathrm{s}} + \nabla
\boldsymbol{v}_{\mathrm{s}}^\mathrm{T}) - \phi_0 \phi
\mathbf{\hat{g}}, \eeq where $\xi = \zeta - \frac{2}{3}\eta$ and
$\mathbf{\hat{g}}$ is a unit vector in the direction of gravity.
Neglecting terms $\textit{O}(\phi_0)$, moving in the reference frame
of the matrix, and introducing constitutive laws $K = \phi^n$ and $\xi
= \phi^{-m}$ for the permeability and a combined matrix viscosity, the
system \eqref{eq: evolS}-\eqref{eq: shear} reduces to the
dimensionless, 1D form for the vertical ascent of the fluid magma
\beq \label{eq: timestepper} \phi_t = \phi^m \mathcal{P}, \eeq
\beq \label{eq: elliptic} -(\phi^n \mathcal{P}_z)_z + \phi^m
\mathcal{P} = -(\phi^n)_z \prd \eeq After normalizing by the natural
length and time scales \eqref{eq: scales}, this system of equations
conveniently has no coefficient dependence on adjustable parameters,
which gives rise to a scaling symmetry discussed in Section \ref{sec:
  ldr_scaling}.  The constitutive power laws represent the expected
effects of changes in the matrix porosity on its permeability and
combined viscosity, when the porosity is small.  Both physical
arguments \citep[c.f.][]{scott84,scott86} and homogenization theory
\citep{simpson10} have been used to argue that physically relevant
values for the constitutive parameters $(n,m)$ lie in the range $n \in
[2,5]$ and $m \in [0,1]$ or at least in some subset of that range.
Eliminating the compaction pressure from the above formulation, leads
to the scalar magma equation \eqref{eq: magma} considered in this
paper.

From the derivation, we observe that the time evolution of porosity in an interpenetrating magma flow system is controlled by a nonlinear advection term $(\phi^n)_z$ and a dispersive term $(\phi^n(\phi^{-m}\phi_t)_z)_z$.  The nonlinearity enters the system via buoyant forcing of the melt, driven by the matrix permeability.  Compaction and dilation of the matrix generate dispersive effects on the melt which for step-like initial data, result in porosity propagation in the form of an expanding region of porosity waves.


\subsection{Viscous Fluid Conduits} \label{sec: Fluid Problem}
An independent formulation of eq. \eqref{eq: magma} arises in the context of a conduit of buoyant fluid ascending through a viscously deformable pipe.  For magma, this represents an alternative transport regime to the interpenetrating flow described above, most closely related to flow up the neck of a thermal plume in the mantle.

Following \citet{olson86}, the buoyant fluid rises along a vertical conduit of infinite length with circular cross-sections, embedded in a more viscous matrix fluid.  The matrix with density $\rho_\mathrm{M}$, viscosity $\eta_\mathrm{M}$ and the fluid conduit with density $\rho_\mathrm{f}$, viscosity $\eta_\mathrm{f}$, must satisfy
\beq
	\rho_\mathrm{M} > \rho_\mathrm{f} \quad \quad \eta_\mathrm{M}\gg\eta_\mathrm{f}	\prd
\eeq
The circular cross-sectional area is $A'=\pi R'^2$, where the conduit radius $R'$ is allowed to vary.  The Reynolds' number and slope of the conduit wall (i.e. the ratio of conduit deformation to wavelength) are assumed to be small and mass and heat diffusion are negligible.

With this set-up, the conduit flux $Q'$  can be related to the
cross-sectional area $A'$ via Poiseuille's Law for pipe flow of a Newtonian fluid (recall
that primed variables are dimensional)
\beq
	Q' = -\frac{A'^2}{8\pi \eta_\mathrm{f}} \frac{\partial P'}{\partial z'}
	\label{eq: poiseuille}
\eeq
where $P$ is the nonhydrostatic pressure of the fluid.  Conservation of mass manifests as
\beq
	\frac{\partial A'}{\partial t} + \frac{\partial Q'}{\partial z} = 0 \prd
	\label{eq: continuity}
    \eeq To derive an expression for $P'$, we balance radial forces at
    the conduit wall.  Using the small-slope approximation, we assume
    radial pressure forces in the conduit dominate viscous effects.
    In the matrix, the radial components of the normal force dominate
    viscous stresses.  Setting the radial forces of the matrix and
    conduit equal at the boundary and making the appropriate
    small-slope approximations, yields an expression for the
    nonhydrostatic pressure \beq P' = -\Delta \rho g z' +
    \frac{\eta_\mathrm{M}}{A'} \frac{\partial A'}{\partial t'},
	\label{eq: pl}
    \eeq for $\Delta \rho = \rho_\mathrm{M} - \rho_\mathrm{f}$.
    Substitution of \eqref{eq: pl} back into \eqref{eq: poiseuille},
    utilizing \eqref{eq: continuity} for simplification and
    nondimensionalizing about a background, steady state, vertically
    uniform Poiseuille flow \beqs Q_0 = \frac{\Delta \rho g}{8 \pi
      \eta_\mathrm{f}} A_0^2, \eeqs with length and time scales $L$ and
    $T$ \beqs L =
    \left(\frac{\eta_\mathrm{M}A_0}{8\pi\eta_\mathrm{f}}\right)^{\frac{1}{2}},
    \quad T = \frac{1}{\Delta \rho g}\left(\frac{8\pi
        \eta_\mathrm{f}\eta_\mathrm{M}}{A_0}\right)^{\frac{1}{2}},
    \eeqs gives the non-dimensional equations \beqs Q =
    A^2\left[1+\frac{\partial}{\partial
        z}\left(\frac{1}{A}\frac{\partial Q}{\partial
          z}\right)\right], \eeqs \beqs -\frac{\partial A}{\partial t} =
    \frac{\partial Q}{\partial z} \prd \eeqs which are eqs. \eqref{eq:
      timestepper} and \eqref{eq: elliptic} for the case $(n,m)=(2,1)$.


\section{Properties of the Magma Equation}\label{sec: Equation Properties}
In this section we recall several results that will be important for
our studies of magma DSWs.  It is interesting to note that for the
pairs $(n,m) = (-1,0)$ and $(n,m) = (\frac{1}{2},\frac{1}{2})$ the
magma equation has been shown to be completely integrable, but for
other rational values of $(n,m)$ it is believed to be non-integrable
by the Painleve ODE test \citep{harris06}.  We will primarily be
concerned with the physically relevant non-integrable range $m\in
[0,1], n\in[2,5]$, but many of the results are generalized to a much
wider range of values.


\subsection{Linear Dispersion Relation and Scaling}	\label{sec: ldr_scaling}
Linearizing eq. \eqref{eq: magma} about a uniform background porosity $\Phi$ and seeking a harmonic solution with real-valued wavenumber $k$, and frequency $\om_0$, we write $\phi$ as 
\beq
	\phi(z,t) \approx \Phi + \nu \lp e^{i(kz - \om_0 t)} + \mathrm{c.c.}  \rp, \quad \abs{\nu} \ll 1,
\eeq
where $\mathrm{c.c.}$ denotes the complex conjugate.  Substitution into eq. \eqref{eq: magma} yields the linear dispersion relation
\beq
	\om_0(k, \Phi) = \frac{n \Phi^{n-1}k}{1+ \Phi^{n-m}k^2}
	\label{eq: lindisp}	\prd
\eeq
Taking a partial derivative in $k$ gives the group velocity
\beq	\label{eq: lingroupvel}
	(\om_0(k,\Phi))_k = \frac{n\Phi^{n-1}(1-\Phi^{n-m}k^2)}{(1+\Phi^{n-m}k^2)^2}	\prd	
\eeq
Note that although the phase velocity $\frac{\om_0}{k}$ is strictly positive, the group velocity \eqref{eq: lingroupvel} can take on either sign, with a change in sign occurring when $k^2 = \Phi^{m-n}$.  Taking a second partial derivative of eq. \eqref{eq: lindisp} with respect to $k$ gives
\beq	\label{eq: signdisp}
	(\om_0(k,\Phi))_{kk} = \frac{-2nk\Phi^{2n-1-m}(3-\Phi^{n-m}k^2)}{(1+\Phi^{n-m}k^2)^3}	\prd
\eeq	
Introducing the sign of dispersion, we say the system has positive dispersion if $(\om_0(k,\Phi))_{kk}>0$ so that the group velocity is larger for shorter waves.  Similarly, negative dispersion is defined as $(\om_0(k,\Phi))_{kk}<0$.  From eq. \eqref{eq: signdisp}, the sign of dispersion is negative for long waves but switches to to positive when $k^2 = 3\Phi^{m-n}$.  These distinguished wavenumbers, the zeros of the group velocity and sign of dispersion, have physical ramifications on magma DSWs that will be elucidated later.

The magma equation also possesses a scaling symmetry.  It is invariant under the change of variables
\beq	\label{eq: symmetry}
	\phi' = \Phi \phi, \quad z' = \Phi^{\frac{n-m}{2}} z, \quad t' = \Phi ^{\frac{1}{2}(2-n-m)} t	, \quad \Phi >0	\prd
\eeq
This allows us to normalize the background porosity to one without loss of generality, which we will do in Section \ref{sec: DSWs}.


\subsection{Long Wavelength Regime}	\label{sec: long wavelength}
In the weakly nonlinear, long wavelength regime, the magma equation reduces to the integrable KdV equation \citep{whitehead86, takahashi90, elperin94} .  To obtain KdV we enter a moving coordinate system with the linear wave speed $n$ and introduce the ``slow'' scaled variables $\zeta = \eps^{1/2}(z-n t)$, $\tau = \eps^{3/2} t$, and $\phi(z,t) = 1+ \eps u(z,t)$ to the magma equation.  Assuming $|\eps| \ll 1$, a standard asymptotic calculation results in 
\beq
	u_\tau + n(n-1)uu_\zeta + nu_{\zeta \zeta \zeta} = 0,
	\label{eq: kdv}
\eeq
which has no dependence on the parameter $m$.  What this implies physically is that dispersion in the small amplitude, weakly nonlinear regime is dominated by matrix compaction and dilation.  Nonlinear dispersive effects resulting from matrix viscosity are negligible.  The original construction of a DSW was undertaken for KdV by \citet{gurevich74}.  In Section \ref{sec: DSWs}, we will compare the results of KdV DSW theory with our results for magma DSWs.


\subsection{Nonlinear Periodic Traveling Wave Solutions}	\label{sec: traveling wave}
The well-studied solitary waves of eq. \eqref{eq: magma} are a limiting case of more general periodic traveling wave solutions.  To apply Whitham theory to the magma equation, it is necessary to derive the periodic traveling wave solution to eq. \eqref{eq: magma}, which forms the basis for nonlinear wave modulation.  In the special case $(n,m)=(3,0)$, \citet{marchant05} obtained an implicit relation for $\phi$ in terms of elliptic functions, and for $(n,m)=(2,1)$, the equation was derived in integral form by \citet{olson86}.  Here we consider the full physical range of the constitutive parameters.

We seek a solution of the form $\phi(z,t)=\phi(\theta)$, where $\theta=z-ct$ for wave velocity $c$ ($c$ is related to the wave frequency $\omega$ and the wavenumber $k$ by the relation $\omega = c k$), such that $\phi(\theta) = \phi(\theta+L)$ with wavelength $L$.  Inserting this ansatz into eq. \eqref{eq: magma} and integrating once yields
\beq \label{eq: 2deriv}
	c(\phi^{-m}\phi')'=A\phi^{-n}+c\phi^{1-n}-1,
\eeq
for integration constant $A$ and $'$ indicating a derivative with respect to $\theta$.  Observing 
\beqs	
	c(\phi^{-m}\phi')' = c\lp\frac{1}{2}\phi^m\rp\left[\frac{d}{d\phi}\lp\phi^{-2m}(\phi')^2\rp\right],
\eeqs
enables us to integrate with respect to $\phi$ and find for $m \ne 1$ and $n+m \ne 2$
\beq \label{eq: potential1}
	(\phi')^2=\frac{2}{2-m-n}\phi^{m-n+2}-\frac{2}{c(1-m)}\phi^{m+1}-\frac{2A}{c(1-m-n)}\phi^{m-n+1} -
		\frac{2B}{c}\phi^{2m},
\eeq
with a second integration constant $B$.  

For $n+m=2$, which for our purposes necessarily implies $n=2$ and $m=0$, we integrate eq. \eqref{eq: 2deriv} to
\beq \label{eq: potential2}
	(\phi')^2= 2\ln{\phi} - \frac{2A}{c}\phi^{-1} - \frac{2}{c}\phi + 
		\frac{2B}{c} \prd
\eeq
For the case $m=1$, $n \in [2,5]$, eq. \eqref{eq: 2deriv} integrates to
\beq \label{eq: potential3}
	(\phi')^2= -\frac{2}{c}\phi^{2}\ln{\phi} + \frac{2A}{cn}\phi^{2-n} + \frac{2}{1-n}\phi^{3-n} + 
		\frac{2B}{c}\phi^{2}	\prd
\eeq

Equations \eqref{eq: potential1}, \eqref{eq: potential2}, and
\eqref{eq: potential3} can be written in the general form $(\phi')^2 =
g(\phi)$, where $g(\phi)$ is the potential function.  Periodic
solutions exist when $g(\phi)$ has three real, positive roots such
that $0<\phi_1<\phi_2<\phi_3$.  In this case, the potential function
can be rewritten \beq \label{eq: potential} (\phi')^2 = g(\phi) =
(\phi_1 - \phi)(\phi_2 - \phi)(\phi_3-\phi) r^2(\phi) \eeq where
$r(\phi)$ is some smooth function and $r(\phi) \ne 0$ for $\phi \in
(0,\phi_3)$.  The sign is chosen so that $(\phi')^2 \xrightarrow{\phi
  \rightarrow \infty} -\infty$ for $c>0$ as in each of the equations
\eqref{eq: potential1}, \eqref{eq: potential2}, and \eqref{eq:
  potential3}.  We verify that $c>0$ because in the linear wave limit
when $\phi_2 \rightarrow \phi_3$, $c=\frac{\om_0}{k}$ which is
positive from eq. \eqref{eq: lindisp}.  Similarly from solitary wave
derivations \eqref{eq: ampspeed}, noting that $A>1$, the speed is
positive.  From eqs. \eqref{eq: potential1}, \eqref{eq: potential2},
\eqref{eq: potential3} and the fact that $c$ depends continuously on
the roots $\phi_1, \phi_2, \phi_3$, the only way for $c$ to pass from
positive to negative would be for $\phi \equiv 0$, else the potential
function is singular, and in neither case does this yield a
non-trivial periodic traveling wave solution.  Thus the wave speed $c$
must be strictly positive.

We also confirm that the potential functions in equations \eqref{eq:
  potential1}, \eqref{eq: potential2}, and \eqref{eq: potential3} have
no more than three positive roots in the physically relevant range of
the constitutive parameters.  First consider the case of integer $n,m$
so that all exponents are integers.  From Descartes' Rule of Signs,
polynomials with real coefficients and terms ordered by increasing
degree have a number of roots at most the number of sign changes in
the coefficients.  Hence, the four-term polynomial expression
\eqref{eq: potential1} has at most three positive roots as we assumed.
For eq.~\eqref{eq: potential2}, taking a derivative of $g(\phi)$
yields a three-term polynomial, which thus has at most two positive
roots.  From the Mean Value Theorem, this means \eqref{eq: potential2}
has no more than three positive roots.  Eq.~\eqref{eq: potential3}
follows in a similar manner, upon noting that one can factor out
$\phi^2$ so that the positive real roots are unchanged.  Now relax the
assumption that the constitutive parameters take on integer values and
suppose they can take on any rational value so that eq.~\eqref{eq:
  potential1} has rational exponents $\{y_1, y_2, y_3, y_4\}$ with
least common multiple $y$.  Then let $\phi = \psi^{y}$.  It follows
that $g(\phi)$ has the same number roots as $g(\psi^y)$, the latter
now a four-term polynomial in $\psi$ with integer exponents.  As
above, this may have at most three positive roots.  A similar argument
follows for eqs.~\eqref{eq: potential2}, \eqref{eq: potential3}.  The
case of irrational $n,m$ follows from continuity and density of the
rationals in the real line.  This gives an upper bound on the number
of positive roots, which must be either one or three.  We will
consider only the cases where the particular parameters give exactly
three positive roots.

\begin{figure}
  \centerline{\includegraphics{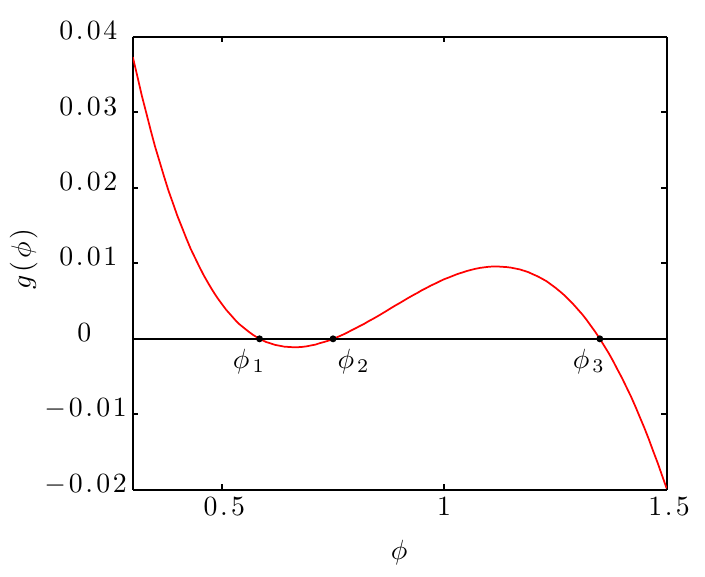}}
  \caption{Plot of $(\phi')^2 = g(\phi)$, $(n,m)=(2,1)$.  The bounded, periodic solution lies between the roots $\phi_2$ and $\phi_3$.  The roots correspond to a particular choice of the physical wave parameters $a=0.6$, $k \approx 0.29$, and $\bar{\phi} = 1$.}
\label{fig: potential}
\end{figure}

One can see from eq. \eqref{eq: potential} that there is a map between
the parameters $A,B,c$ and the roots $\phi_1, \phi_2,\phi_3$ so the
periodic traveling wave solution can be determined by use of either
set of variables.  There is further an additional set of ``physical''
variables which we will use here.  They are the periodic wave
amplitude $a$ \beq \label{eq: amplitude} a = \phi_3 - \phi_2, \eeq the
wavenumber $k = \frac{2\pi}{L}$, which we express \beq \label{eq:
  wavenumber} k = \frac{2\pi}{L} = \pi \left(\int_{\phi_2}^{\phi_3}
  \frac{1}{\sqrt{g(\phi)}} d\phi\right)^{-1}, \eeq and the
wavelength-averaged porosity $\bar{\phi}$, \beq \label{eq: phi_bar}
\bar{\phi} = \frac{2}{L}\int_{\phi_2}^{\phi_3}
\frac{\phi}{\sqrt{g(\phi)}} d\phi \prd \eeq The three parameter family
of periodic waves will be parameterized by either $(\phi_1, \phi_2,
\phi_3)$ or $(a,k,\bar{\phi})$.  Invertibility of the map between
these two parameter families is difficult to verify in general.
However, in the limiting case $\phi_2 \rightarrow \phi_3$, we can
express the physical variables in the following form at leading order:
\beqs a = 0, \quad \bar{\phi} = \phi_2, \quad k =
\frac{r(\phi_2)\pi}{2} \lp \phi_2 - \phi_1 \rp ^\frac{1}{2}, \quad
\phi_1 < \phi_2 \prd \eeqs Hence, the determinant of the Jacobian of
the transformation is $\partial k/\partial \phi_1 \ne 0$.  A similar
argument holds for the solitary wave limiting case using conjugate
variables which will be introduced in Section \ref{sec: Leading Edge}.
Since invertibility holds in the limiting cases, we make the
additional assumption that it holds in general.  Figure \ref{fig:
  potential} shows a plot of $g(\phi)$ for $(n,m)=(2,1)$, with the
particular wave parameters $a=0.6$, $\bar{\phi} = 1$, $k \approx
0.29$.  From the example in figure \ref{fig: potential}, we observe
that the periodic wave takes its values on the bounded, positive
portion of $g(\phi)$ with $\phi_2 \le \phi \le \phi_3$.

The solitary wave solution of the magma equation occurs in the limit $\phi_2 \rightarrow \phi_1$.  To derive the magma solitary wave, we impose the conditions on the potential function $g(\Phi) = 0, g(A) = 0, g^{\prime}(\Phi) = 0$ for a solitary wave of height $A$ which propagates on a positive background value $\Phi$.  Utilizing these conditions on the background state $\Phi = 1$ yields a solitary wave amplitude-speed relation $c(A)$ derived in \citet{nakayama92} for $n,m$ in the physical range
\beq	\label{eq: ampspeed}
\begin{cases}
	\displaystyle
	c(A)=\frac{A^2-2A+1}{A\ln{A}-A+1}, & \mbox{if } m=0, n=2,	 \\[3 mm]	
	\displaystyle
	c(A)=\frac{(n-1)(nA^n\ln{A}-A^n+1)}
		{A^n-nA+n-1}, &
		\mbox{if } m=1, \\[3 mm]		
	\displaystyle
	c(A)=\frac{(n+m-2)[nA^{n+m-1}-(n+m-1)A^n+m-1]}
	{(m-1)[A^{n+m-1}-(n+m-1)A+n+m-2]}, &
		\mbox{otherwise } \prd 
\end{cases}
\eeq

\subsection{Conservation Laws}	\label{sec: conservation}
Another feature of the magma equation necessary for the application of DSW modulation theory is the existence of at least two conservation laws.  \citet{harris96} found two independent conservation laws for the magma equation given arbitrary values for the parameters $(n,m)$.  While there are additional conservation laws for particular values of $(n,m)$, these lie outside the physically relevant range and will not be addressed.  The first conservation law is the magma equation itself written in conservative form.  The second conservation law does not have a clear physical meaning as it does not correspond to a momentum or an energy.  It also possesses singular terms in the dry limit $\phi \rightarrow 0$, but this is not a problem in the positive porosity regime considered here.  In conservative form, the conservation laws for the physically relevant parameter values are 
\beq	\label{eq: conservation1}
	\phi_t + (\phi^n(1-\phi^{-m}\phi_t)_z)_z = 0
\eeq
and
\beq	\begin{cases}	\label{eq: conservation2}
	\displaystyle
	\left(-\frac{1}{2}\phi_z^2 + \ln{\phi}\right)_t + 
		\left(- \phi \phi_{tz}+2\phi\right)_z=0, & \mbox{if } (n,m)=(0,2) \\[6 mm]	
	\displaystyle
	\left(-\frac{n}{2}\phi^{-2}\phi_z^2 + \frac{\phi^{1-n}-1}{1-n}\right)_t + 
		\left(\phi^{-2}\phi_z\phi_t - \phi^{-1}\phi_{tz}+n\ln{\phi}\right)_z = 0, &
		\mbox{if } m=1 \\[6 mm]
	\displaystyle	
	\left(\frac{1-m-n}{2}\phi^{-2m}\phi_z^2 + \frac{1}{2-n-m}\left(\phi^{2-n-m}-1\right)\right)_t + \\ \quad
		\left(m\phi^{-2m}\phi_z\phi_t - \phi^{1-2m}\phi_{tz} + \frac{n}{1-m}\phi^{1-m}\right)_z = 0, &
		\mbox{otherwise.}
\end{cases}
\eeq

\section{Resolution of an Initial Discontinuity}\label{sec: DSWs}

A canonical problem of fundamental physical interest in nonlinear media is the evolution of an initial discontinuity.  In the case of an interpenetrating magma flow, one can imagine magma flowing from a chamber of uniform porosity into an overlying region of lower porosity.  We study DSWs arising from this general set-up.  

Consider the magma equation \eqref{eq: magma} given initial data \eqref{eq: ic_gen} with the requirement $0<\phi_+<\phi_-$, which ensures singularity formation in the dispersionless limit $\phi_t + (\phi^n)_z =0$ (gradient catastrophe).  This is a dispersive Riemann problem.  From DSW theory, a discontinuity in a dissipationless, dispersively regularized hyperbolic system will result in smooth upstream and downstream states connected by an expanding, rapidly oscillating region of rank-ordered, nonlinear waves.  In the negative dispersion regime, this region is characterized by the formation of a solitary wave at the right edge and a linear wave packet at the left.  Our goal in this section is to derive analytical predictions for the speeds at which the edges propagate and the height of the leading edge solitary wave.

Following the general construction of \citet{el05}, which extends the
work of \citet{whitham65} and then \citet{gurevich74} to
non-integrable systems, we will assume that the spatial domain can be
split into three characteristic subdomains,
$\left(-\infty,z_1(t)\right)$, $\left[z_1(t),z_2(t)\right]$, and
$\left(z_2(t),\infty \right)$, where $z_1(t) \le z_2(t)$, for all
$t>0$.  We formally introduce the slow variables $Z=\eps z$ and
$T=\eps t$, where $0<\eps\ll1$, which is equivalent to considering $t
\gg 1$ when $T=\textit{O}(1)$.  For the analysis, we use the slow
variables $Z,T$ to describe the modulations of the traveling wave.
For the numerics, we consider $t\gg 1$ and do not rescale.  Inside the
dispersive shock region $z\in[z_1(t),z_2(t)]$, the solution is assumed
to be described by slow modulations of the nonlinear, single phase,
periodic traveling wave solution, i.e. a modulated one-phase region
(the use of phase here describes the phase of the wave, not to be
confused with the physical phase of the flow, e.g. melt or matrix).  The
modulation equations are a system of quasi-linear, first-order PDE
formed by averaging over the two conservation laws augmented with the
conservation of waves equation.  This allows for the description of
the DSW in terms of the three slowly varying, physical wave variables:
the wavenumber $k$ \eqref{eq: wavenumber}, the average porosity
$\bar{\phi}$ \eqref{eq: phi_bar}, and the periodic wave amplitude $a$
\eqref{eq: amplitude}.  Outside the dispersive shock, the dynamics are
slowly varying so the solution is assumed to be governed by the
dispersionless limit of the full equation which we call the zero-phase
equation.  The dynamic boundaries between the zero- and one-phase
regions are determined by employing the Gurevich-Pitaevskii matching
conditions \citep{gurevich74}.  The behavior of the modulation system
near the boundaries allows for a reduction of the modulation system of
three quasilinear PDE to only two, thus locally reducing the
complexity of the problem.  For a simple wave solution, the limiting
modulation systems can be integrated uniquely along an associated
characteristic with initial data coming from the behavior at the
opposite edge.

This construction relies on several assumptions about the mathematical structure of the magma equation and its modulation equations.  From our discussion in Section \ref{sec: Equation Properties}, it is immediate that we have a real linear dispersion relation with hyperbolic dispersionless limit (seen formally by substituting the slow variables defined above and taking only the leading order eqation).  The two conservation laws given in equations \eqref{eq: conservation1} and \eqref{eq: conservation2} and the existence of a nonlinear periodic traveling wave with harmonic and solitary wave limits round out the basic requirements.  Beyond these, we must satisfy additional constraints on the modulation system.  Its three characteristics are assumed real and distinct so that the modulation equations are strictly hyperbolic and modulationally stable.  The final requirement is the ability to connect the edge states by a simple wave.  For this, the two states are connected by an integral curve associated with only one characteristic family.  More will be said about this condition in the following discussion.  We now supply the details of the application of El's method to the dispersive Riemann problem for the magma equation.

\subsection{Application of El's method}	\label{sec: method}
To ``fit'' a modulated dispersive shock solution, we consider first the solution within the DSW, $z \in [z_1(t), z_2(t)]$.  In this region, the solution is described locally by its nonlinear periodic traveling wave solution, parameterized by its three roots $\phi_1, \phi_2, \phi_3$, which vary slowly, depending on $Z, T$.  Variations in the periodic traveling wave solution are governed by averaging over the system of conservation equations \eqref{eq: conservation1} and \eqref{eq: conservation2}, along with an additional modulation equation known as conservation of waves,  
\beq
	k_T +\om_Z = 0,
\eeq
where $k$ is the wavenumber and $\om$ is the nonlinear wave frequency \citep{whitham65}.  Recalling the wavenumber $k$ \eqref{eq: wavenumber}, we define the wavelength average of a generic smooth function $F(\phi)$ to be
\beq
	\overline{F(\phi)}=\frac{k}{\pi}\int_{\phi_2}^{\phi_3}\frac{F(\phi)}{\sqrt{g(\phi)}}d\phi	\prd
\eeq
It is convenient here to use the nonlinear wave parameterization in terms of the three roots of $g(\phi)$.  Later we will use the physical wave variables $a, k, \bar{\phi}$.

We now seek moving boundary conditions for $z_1(t)$ and $z_2(t)$ where the modulation solution is matched to the dispersionless limit.  For the magma equation, the dispersionless limit is the zero-phase equation
\beq	\label{eq: zero phase}
	\tilde{\phi}_T + n \tilde{\phi}^{n-1}\tilde{\phi}_Z = 0	\prd
\eeq
The GP boundary conditions \citep{gurevich74} include the matching of the average value of the porosity to the dispersionless solution $\bar{\phi} \rightarrow \tilde{\phi}$, as well as a condition on the amplitude and wavenumber.  Recalling figure \ref{fig: schematic}, at the trailing edge, the wave amplitude vanishes.  At the leading edge, the leading wave is assumed to take the form of a solitary wave and thus the wavenumber decays to zero.  This could also be stated in terms of the roots of the potential function, as we saw in Section \ref{sec: traveling wave} where the diminishing amplitude limit corresponds to $\phi_2 \rightarrow \phi_3$ and the solitary wave limit to $\phi_2 \rightarrow \phi_1$.  This implies that the boundary conditions align with double roots of the potential function.  This orientation of the DSW with a rightmost solitary wave and a leftmost linear wave packet is expected for systems with negative dispersion, as is the case here, for at least small jumps \citep{el05}.  These conditions result in the following moving boundary conditions for an initial discontinuity:
\begin{align}
	&z=z_1(t): \quad \overline{\phi}=\phi_-,  \quad a=0, \quad \phi_2 \rightarrow \phi_3,\\
	&z=z_2(t): \quad \overline{\phi}=\phi_+, \quad k=0,	\quad \phi_2, \rightarrow \phi_1 \prd
\end{align}
In the $a \rightarrow 0$ and $k\rightarrow 0$ limits, one can show
that $\overline{F(\phi)} = F(\overline{\phi})$.  This is a fundamental
mathematical property of averaging that allows for the Whitham system
of three modulation equations to be reduced exactly to a system of two
equations at the leading and trailing edges $z_1(t), z_2(t)$.  An
important note is that in the limit of vanishing amplitude, the
nonlinear wave frequency $\om = \om(k,\bar{\phi},a)$ becomes the
linear dispersion relation $\om_0(k,\bar{\phi})=\om(k,\bar{\phi},0)$
from eq. \eqref{eq: lindisp} which has no dependence on the amplitude.
The exact reduction of the full modulation system to two PDE enables
the construction of a self-similar simple wave solution.  This simple
wave of each boundary system is directly related to the simple wave
associated with the second characteristic family of the full Whitham
modulation system via the integral curve connecting the left state
$(\phi_-, k_-, 0)$ to the right state $(\phi_+,0,a_+)$.  The goal is
to determine $k_-$ and $a_+$ and evaluate the characteristic speeds at
the edges.  Given $\phi_-$ and $\phi_+$, this construction provides
the four key physical properties of the magma DSW as in figure
\ref{fig: schematic}.  In the next two sections, we implement this
method.

\subsection{Determination of the Trailing Edge Speed}	\label{sec: Trailing Edge}

To determine the trailing edge speed of the DSW, we consider the above modulation system in a neighborhood of $z_1(t)$, where implementation of the boundary conditions yields the reduced limiting system
\beq \label{eq: redsys1}
	\bar{\phi}_T + (\bar{\phi}^n)_Z = 0,
\eeq
\beq \label{eq: redsys2}
	k_T + (\om_0)_Z = 0	\prd
\eeq
In characteristic form, the above system reduces to
\beq	\label{eq: trailchar}
	-(\om_0)_{\bar{\phi}} \frac{\ud \bar{\phi}}{\ud T} + \left(n \bar{\phi}^{n-1} - (\om_0)_k\right) \frac{\ud k}{\ud T} = 0
\eeq
along the characteristic curve $\frac{\ud Z}{\ud T} = (\om_0)_k$.  The characteristic speed is the linear wave group velocity \eqref{eq: lingroupvel} but here with $\Phi \rightarrow \bar{\phi}$.  To determine the linear wave speed, we integrate eq. \eqref{eq: trailchar} along the characteristic by introducing $k=k(\bar{\phi})$.  Integrating from the leading edge solitary wave where the wavenumber vanishes, i.e. $k(\phi_+)$=0, back across the DSW to the trailing edge determines the linear wavenumber $k_-$.  Put another way, we connect states of our modulation system at the trailing edge $(k,\bar{\phi}) = (k_-,\phi_-)$ to the leading edge in the $a=0$ plane $(k,\bar{\phi}) = (0,\phi_+)$, by assuming $k$ varies in only one characteristic family.  This is our simple wave assumption.  To determine the speed, we evaluate $s_-=(\om_0)_k$ at $(k_-, \phi_-)$.  The fact that we can restrict to the $a=0$ plane follows from the assumed existence of an integral curve for the full Whitham modulation system and the GP matching conditions at the solitary wave edge, which are independent of $a$.  Recalling the symmetry \eqref{eq: symmetry}, without loss of generality, we restrict to the case
\beq \label{eq: IC}
	\phi(z,0) = 
	\begin{cases}
		1, & z \in (-\infty,0] \\
		\phi_+, & z \in (0,\infty),
	\end{cases}
\eeq
where $\phi_+ \in (0,1)$.  From the physical derivation, eq. \eqref{eq: magma} has already been non-dimensionalized on a background porosity scale so it is natural to normalize the upstream flow to unity in the dimensionless problem.  Restriction of $\phi_+ \in (0,1)$ results from physical interest in the problem of vertical flow from a magma chamber into a dryer region above, but one through which magma may still flow.

To determine the linear edge wavenumber $k_-$, it is necessary to solve the ordinary differential equation (ODE) initial value problem (IVP) resulting from the simple wave assumption $k=k(\bar{\phi})$ in eq. \eqref{eq: trailchar}
\beq	\label{eq: kint} 
	\frac{\ud k}{\ud\bar{\phi}}= \frac{(\om_0)_{\bar{\phi}}}{n\bar{\phi}^{n-1} - (\om_0)_k} = 
	\frac{k\left[n-1-(1+m)k^2\bar{\phi}^{n-m}\right]}{\bar{\phi}\left[(1+\bar{\phi}^{n-m}k^2)^2 - 
		(1-\bar{\phi}^{n-m}k^2) \right]}, \quad k(\phi_+)=0	\prd
\eeq
To find an integral of \eqref{eq: kint}, it is convenient to use the change of variables
\beq
	\alpha = \frac{\om_0}{c(\bar{\phi})k}, \quad c(\bar{\phi}) = n\bar{\phi}^{n-1}, 
\eeq
which, upon substitution, is
\beq	\label{eq: alphadef}
	\alpha = \frac{1}{1+\bar{\phi}^{n-m}k^2}	\prd
\eeq
The IVP \eqref{eq: kint} therefore becomes
\beq	\label{eq: alphaint}
	\frac{\ud\alpha}{\ud\bar{\phi}}=-\frac{\left[(m+n-2)\alpha+n-m\right]\alpha}{\bar{\phi}(2\alpha+1)},	
	\quad \alpha(\phi_+)=1 \prd
\eeq
Equation \eqref{eq: alphaint} is separable with an integral giving an implicit relation between $\alpha$ and $\bar{\phi}$.  Defining $M = \frac{n-m}{n+m-2}$, the relation is
\beq	\label{eq: implicitalpha}
\begin{cases}
	\displaystyle
	\alpha e^{2(\alpha-1)} = \left(\frac{\bar{\phi}}{\phi_+}\right)^{2-2n}, & m+n=2 \\[6mm]
	\displaystyle
	\alpha \left(\frac{\alpha + M}{1+M}\right)^{2M-1} = \left(\frac{\bar{\phi}}{\phi_+}\right)^{m-n}, & 	\text{otherwise}
	\prd
\end{cases}
\eeq
The implicit function theorem proves that eq. \eqref{eq: implicitalpha} can be solved for $\alpha(\bar{\phi})$, provided $\alpha \ne -\frac{1}{2}$.  From eq. \eqref{eq: alphaint}, $\alpha = -\frac{1}{2}$ corresponds to a singularity in the righthand side of the ODE.  Moreover, negative values of $\alpha$ lead to negative average porosity which is unphysical.  Therefore, we can generally solve eq. \eqref{eq: implicitalpha} to get $\alpha = \alpha(\bar{\phi})$.

To find the speed of the trailing linear wave packet $s_-$, we evaluate eq. \eqref{eq: implicitalpha} at the trailing edge.  There we find $\alpha(1) = \alpha_-$, with the initial condition \eqref{eq: IC}, satisfies the implicit relation
\beq	\label{eq: implicitalphatrail}
\begin{cases}
	\displaystyle
	\alpha_- e^{2(\alpha_--1)} =\phi_+^{2(n-1)}, & m+n=2, \\[3mm]
	\displaystyle
	\alpha_- \left(\frac{\alpha_- + M}{1+M}\right)^{2M-1} = \phi_+^{n-m}, & \text{otherwise}
	\prd
\end{cases}
\eeq
Given particular values for the parameters, we can solve this expression for $\alpha_-$, analytically or numerically.  Note that for the physical range of $(n,m)$, $\alpha_-$ is a decreasing function of $\phi_+$.  Via the transformation \eqref{eq: alphadef}, we obtain $k_-$, the wavenumber at the trailing edge.  Upon substitution into the group velocity \eqref{eq: lingroupvel}, the trailing edge speed is
\beq \label{eq: linspeed}
	s_-=	\frac{-n\left(k_-^2-1\right)}{\left(k_-^2 + 1\right)^2} = -n\alpha_-(1-2\alpha_-)	\prd
\eeq

In general, we cannot find an explicit relation for $\alpha_-$ in \eqref{eq: implicitalphatrail} analytically, but when $m=1$, $M=1$ and the expression simplifies to a quadratic equation in $\alpha_-$ whose physical solution is
\beq	\label{eq: alphaM1}
	\alpha_- = -\frac{1}{2} \left[ 1-\left( 1+8 \phi_+^{n-1} \right) ^ \frac{1}{2} \right], \quad
	m=1	\prd
\eeq
Then for $m=1$, substitution of eq. \eqref{eq: alphaM1} into eq. \eqref{eq: linspeed} yields the trailing edge speed
\beq	\label{eq: linspeedM1special}
	s_- = \frac{1}{2}n \left[3+8\phi_+^{n-1}-3\left(1+8\phi_+^{n-1}\right)^{\frac{1}{2}} \right], \quad
	m=1	\prd
\eeq
Equation \eqref{eq: linspeedM1special} provides a simple formula for the trailing edge speed in terms of the parameter $n$ and the jump height $\phi_+$ when $m=1$, e.g. in viscous fluid conduits.

An interesting physical question to consider is whether the trailing
edge speed can take on negative values for some choice of the
parameters.  Recall that even though the phase velocity is always
positive, the linear group velocity \eqref{eq: lingroupvel},
corresponding to the trailing edge speed, can be negative. Returning
to the problem of vertical magma flow from a chamber, such a result
would imply that for a magma chamber supplying a matrix of
sufficiently small porosity relative to the chamber, porosity waves
could transmit back into the chamber and cause the matrix within the
chamber to compact and distend.  We refer to this condition as
backflow.  From our discussion in Section \ref{sec: ldr_scaling}, the
group velocity evaluated at the trailing edge becomes negative when
$k_-^2 = \phi_-^{m-n}$, or $\alpha_- = \frac{1}{2}$.  Using the
determination of $\alpha_-$ \eqref{eq: implicitalphatrail}, backflow
occurs for any choice of $(n,m)$ when $\alpha_- \le
\frac{1}{2}$. Substituting this value into eq. \eqref{eq:
  implicitalphatrail} gives the critical value $\phi_+ =
\phi_\mathrm{b}$ such that for any $\phi_+ < \phi_\mathrm{b}$, $s_- <
0$: \beqs \label{eq: phi_b} \phi_\mathrm{b} =
\begin{cases}
	\displaystyle
	 \left( \frac{1}{2 e} \right)^{\frac{1}{2(n-1)}},  & m+n=2,	\\[3mm]
	\displaystyle
	 \left\{\frac{1}{2} \left[\frac{1 + 2M}{2(1+M)}\right]^{2M-1}\right\}^{\frac{1}{n-m}}, &
	\text{otherwise}\prd
\end{cases}
\eeqs
Negative propagation of porosity waves for sufficiently large jumps was observed numerically in \citet{spiegelman93b} but could not be explained using viscous shock theory.  Here we have identified an exact jump criterion that initiates backflow.


\subsection{Determination of the Leading Edge Speed and Amplitude}	\label{sec: Leading Edge}

The leading edge speed could be derived in a similar fashion to the
trailing edge, but \citet{el05} describes a simpler approach by
introducing a different system of basis modulation variables. The main
idea is to mirror the description of the linear wave edge by
introducing conjugate variables so that the potential curve $g(\phi)$,
as in figure \ref{fig: potential}, is reflected about the $\phi$ axis.
Then, averaging is carried out by integrating over the interval
$(\phi_1,\phi_2)$ where $-g(\phi) > 0$.  The $\phi_2 \to \phi_1$ limit
at the soliton edge now resembles the $\phi_2 \to \phi_3$ limit at the
linear edge. The conjugate wavenumber $\tilde{k}$ is defined as \beq
\tilde{k}=\pi \lp
\int_{\phi_1}^{\phi_2}\frac{d\phi}{\sqrt{-g(\phi)}}\rp^{-1}, \eeq and
will play the role of the amplitude $a$.
$\Lambda=\frac{k}{\tilde{k}}$ will be used instead of the wavenumber
$k$.  In the conjugate variables, the asymptotic matching conditions
become
\begin{align*}	
	&z=z_1(t): \quad \overline{\phi}=1,  \quad \tilde{k}=0, \quad \phi_2 \rightarrow \phi_3,\\
	&z=z_2(t): \quad \overline{\phi}=\phi_+, \quad \Lambda=0,\quad \phi_2 \rightarrow \phi_1 \prd
\end{align*}
In the modulation system, the reduction of the magma conservation laws to the dispersionless limit  \eqref{eq: zero phase} is retained, but the conservation of waves condition is rewritten in the new variables.  To do so, it is helpful to define the conjugate wave frequency $\tilde{\om}=\tilde{\om}(\tilde{k}, \bar{\phi},\Lambda) = -i\om(i\tilde{k}, \bar{\phi},a(\tilde{k},\Lambda))$.  Assuming the existence of a simple wave, the $\Lambda \rightarrow 0$ limiting behavior of the modulation system takes the form 
\beq	\label{eq: k_int}
	\frac{\ud\tilde{k}}{\ud\bar{\phi}} = 
		\frac{ \tilde{\om}_{\bar{\phi}}}{n\bar{\phi}^{n-1} - \tilde{\om}_{\tilde{k}}},
\eeq
and
\beqs	
	\Lambda_T + \frac{\tilde{\om}}{\tilde{k}}\Lambda_Z =
    \text{$\textit{O}$}(\Lambda \Lambda_Z) 
		\prd
\eeqs
Then the integral curve $\tilde{k} = \tilde{k}(\bar{\phi})$ satisfying eq. \eqref{eq: k_int}, which is the same expression as in the leading edge system \eqref{eq: kint} but in the conjugate variables, corresponds to the characteristics $\frac{dZ}{dT} = \frac{\tilde{\om}_\mathrm{s}(\tilde{k},\bar{\phi})}{\tilde{k}}$, where now the characteristic speed is the conjugate phase velocity. Here $\tilde{\om}_\mathrm{s}(\tilde{k},\bar{\phi})=\tilde{\om}(\tilde{k},\bar{\phi},0)$ is called the solitary wave dispersion relation and is obtained from its expression in terms of the linear dispersion relation $\tilde{\om}_\mathrm{s}(\tilde{k},\bar{\phi})=-i\om_0(i\tilde{k},\bar{\phi})$.  From eq. \eqref{eq: lindisp}, the solitary wave dispersion relation is
\beq
	\tilde{\om}_\mathrm{s}(\bar{\phi},\tilde{k})=\frac{n \bar{\phi}^{n-1}\tilde{k}}{1- \bar{\phi}^{n-m}\tilde{k}^2}
	\prd
\eeq
Upon substitution into eq. \eqref{eq: k_int} and recalling that $\tilde{k}$ behaves like an amplitude, the GP matching condition at the trailing edge takes the form $\tilde{k}(1)=0$ so that we arrive at the IVP
\beq
	\frac{\ud\tilde{k}}{\ud\bar{\phi}}=\frac{\tilde{k}\left[n-1-(1+m)\tilde{k}^2\bar{\phi}^{n-m}\right]}{\bar{\phi}\left[(1+\bar{\phi}^{n-m}\tilde{k}^2)^2 - 
		(1-\bar{\phi}^{n-m}\tilde{k}^2) \right]}, \quad
	\tilde{k}(1)=0	\prd
\eeq
Again, a change of variables will lead to a separable ODE which yields an implicit representation, this time for the conjugate wavenumber at the leading edge.  Defining
\beq
	\tilde{\alpha} = \frac{\tilde{\om}_\mathrm{s}}{c(\bar{\phi})\tilde{k}} = \frac{1}{1-\bar{\phi}^{n-m}\tilde{k}^2},
\eeq
the $\tilde{\alpha}$ IVP is
\beq	\label{eq: betaIVP}
	\frac{\ud\tilde{\alpha}}{\ud\bar{\phi}}=-\frac{\left[(m+n-2)\tilde{\alpha}+n-m\right]\tilde{\alpha}}{\bar		{\phi}(2\tilde{\alpha}+1)},
	\quad \tilde{\alpha}(1)=1 \prd
\eeq
This equation is the same as eq. \eqref{eq: alphaint} but with a different initial location because integration takes place from the trailing edge to the leading edge.  Note that for the physical range of the constitutive parameters $(n,m)$, $\tilde{\alpha}$ is a decreasing function of $\bar{\phi}$.  Integrating \eqref{eq: betaIVP} gives the relation between $\tilde{\alpha}$ and $\bar{\phi}$
\beq	\label{eq: betaimplicit}
\begin{cases}
	\displaystyle
	\tilde{\alpha} e^{2(\tilde{\alpha}-1)} = \bar{\phi}^{2-2n}, & m+n=2 \\[3mm]
	\displaystyle
	\tilde{\alpha} \left(\frac{\tilde{\alpha} + M}{1+M}\right)^{2M-1} = \bar{\phi}^{m-n}, & 	\text{otherwise}
	\prd	
\end{cases}
\eeq
and the requirement $\tilde{\alpha}>-\frac{1}{2}$ for finding $\tilde{\alpha} = \tilde{\alpha}(\bar{\phi})$.  For all physically relevant $\bar{\phi}>0$, $\tilde{\alpha}>0$.  Using the solitary wave dispersion relation, the leading edge speed $s_+$ becomes
\beq \label{eq: solspeed}
	s_+=\frac{\tilde{\om}_\mathrm{s}(\phi_+,\tilde{k}_+)}{\tilde{k}_+} = 
		\frac{n \phi_+^{n-1}}{1- \phi_+^{n-m}\tilde{k}_+^2},
\eeq
 where $\tilde{k}_+ = \tilde{k}(\phi_+)$.  Defining $\tilde{\alpha}_+ = \tilde{\alpha}(\phi_+)$, we find $\phi_+^{n-m}\tilde{k}_+^2 = 1-\frac{1}{\tilde{\alpha}_+}$, which upon substitution into eq. \eqref{eq: solspeed} yields
\beq	\label{eq: solspeed2}
	s_+ = n\phi_+^{n-1}\tilde{\alpha}_+,
\eeq
for any choice of $m$.  To solve for the solitary wave speed, we evaluate the implicit relation for $\tilde{\alpha}$ at the leading edge and insert into \eqref{eq: solspeed2}.  The defining relations for $\tilde{\alpha}_+$ are found by evaluating $\tilde{\alpha}(\phi_+)$ in \eqref{eq: betaimplicit}
\beq \label{eq: betaimplicitlead}
\begin{cases}
	\displaystyle
	\tilde{\alpha}_+ e^{2(\tilde{\alpha}_+-1)} = \phi_+^{2-2n}, & m+n=2 \\[3mm]
	\displaystyle
	\tilde{\alpha}_+ \left(\frac{\tilde{\alpha}_+ + M}{1+M}\right)^{2M-1} =\phi_+^{m-n}, & 	\text{otherwise}
	\prd	
\end{cases}
\eeq
Because $\tilde{\alpha}$ is a decreasing function and $\tilde{\alpha}(1)=1$ and $\phi_+ < 1$, $\tilde{\alpha}_+ = \tilde{\alpha}(\phi_+)$ is an increasing function of $\phi_+$.  As in the trailing edge, the case $m=1$ can be solved explicitly so the leading edge speed is
\beqs 
	s_+ = \frac{1}{2}n
    \left[\left(1+8\phi_+^{n-1}\right)^{\frac{1}{2}} - 1 \right],
    \quad 
	m=1	\prd
\eeqs
Here we have determined an explicit relation between the jump height, the constitutive parameter $n$, and the leading edge speed.

Determination of the leading edge amplitude $a_+$ follows from the
solitary wave amplitude-speed relation \eqref{eq: ampspeed}, rescaled
to a background porosity $\phi_+$ using the scaling symmetry in
eq. \eqref{eq: symmetry}.  In practice, $a_+$ is computed numerically.


\subsection{Analysis of the Theoretical Predictions}		\label{sec: Theoretical}
The theoretical predictions of Sections \ref{sec: Trailing Edge} and \ref{sec: Leading Edge} were limited to explicit speed formulae for the $m=1$ case.  In this section, we extend those results to the full physical range of the constitutive parameters.  We will also check that the speeds satisfy the DSW admissibility criteria \citep{el05} and are consistent with KdV asymptotics in the weakly nonlinear limit.

First, we consider the DSW admissibility criteria.  The dispersionless
limit of the magma equation \eqref{eq: zero phase} has the
characteristic speed $c = n\phi^{n-1}$.  As the DSW evolves, it
continuously expands with speeds $s_-$ at the trailing edge and $s_+$
at the leading edge.  For the DSW construction to be valid, the
external characteristics must impinge upon the DSW region,
transferring information from the dispersionless region into the
modulated one-phase region.  Then the DSW and external characteristic
speeds must satisfy the relations $s_- < n$ and $s_+>n\phi_+^{n-1}$.
These conditions ensure a compressive DSW.  Using the expressions for
the speeds \eqref{eq: linspeed} and \eqref{eq: solspeed2} we find that
in the interior of the shock region \beq s_- = -n\alpha_-(1-2\alpha_-)
< n = c(1), \quad s_+ = n\phi_+^{n-1}\tilde{\alpha}_+ > n\phi_+^{n-1}
= c(\phi_+), \eeq because $\alpha_-<1$ and $\tilde{\alpha}_+>1$ for
$\phi_+ < 1$ as shown earlier.  To admit a solitary wave-led
dispersive shock solution as we have constructed, it must also be the
case that $s_- < s_+$ so that the interior region continues to expand
in time.  We have verified numerically that this condition is
satisfied for choices of the constitutive parameters and the jump
height in the physical range.  Hence, our analytical predictions for
the speeds satisfy the DSW admissibility criteria.

Following our discussion in Section \ref{sec: long wavelength} of the
weakly nonlinear limit $0 < 1-\phi_+ \ll 1$, eqs. \eqref{eq:
  linspeed} and \eqref{eq: solspeed2} must be consistent with the
standard KdV speeds and amplitude for the KdV reduction of the magma
equation \eqref{eq: kdv} \beq \label{eq: kdvtrail} s_- =
n-(1-\phi_+)2n(n-1) \eeq \beq \label{eq: kdvwavenumber} k_- = \left (
  (1 - \phi_+) \frac{2}{3} n (n - 1) \right )^{1/2} \eeq
\beq \label{eq: kdvlead} s_+ = n-(1-\phi_+)\frac{1}{3}n(n-1) \eeq
\beq \label{eq: kdvamp} a_+= 2(1-\phi_+), \quad 0<1-\phi_+ \ll 1 \prd
\eeq Using an asymptotic expansion for $\alpha$ and $\tilde{\alpha}$
near 1 in expressions \eqref{eq: implicitalphatrail} and \eqref{eq:
  betaimplicitlead}, respectively and a small amplitude expansion for
the leading edge wave amplitude $a_+$ in the solitary wave
amplitude-speed relation \eqref{eq: ampspeed}, we have verified that
equations \eqref{eq: kdvtrail} -- \eqref{eq: kdvamp} indeed do
describe the first order asymptotics of the magma results in equations
\eqref{eq: linspeed}, \eqref{eq: solspeed2}, and \eqref{eq: ampspeed}.

\begin{figure}
	\centering
        \subfigure[DSW Speeds, $m=0$]{
        		\label{fig: speedsM0}
              \includegraphics{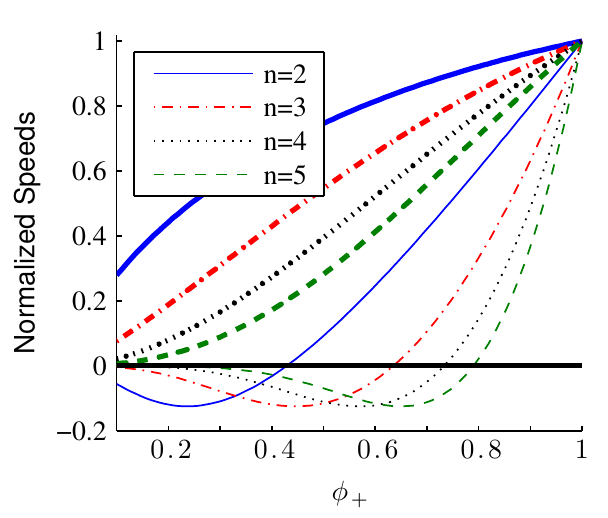}}
	\subfigure[DSW Speeds, $m=1$]{
	\label{fig: speedsM1}
                \includegraphics{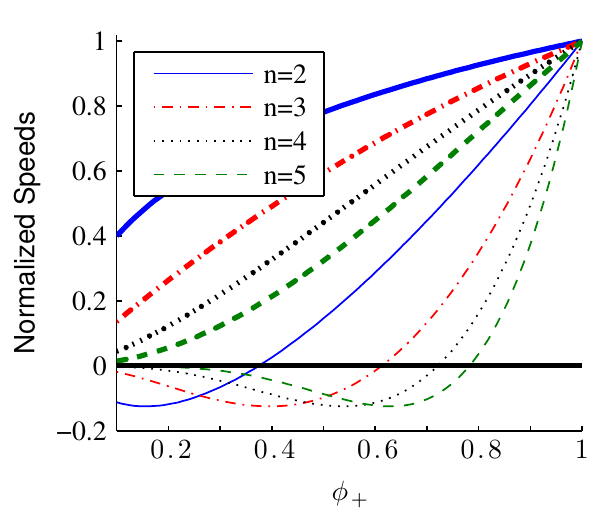}} \\
        \subfigure[Solitary Wave Amplitudes, $m=0$]{
        \label{fig: ampsM0}
                \includegraphics{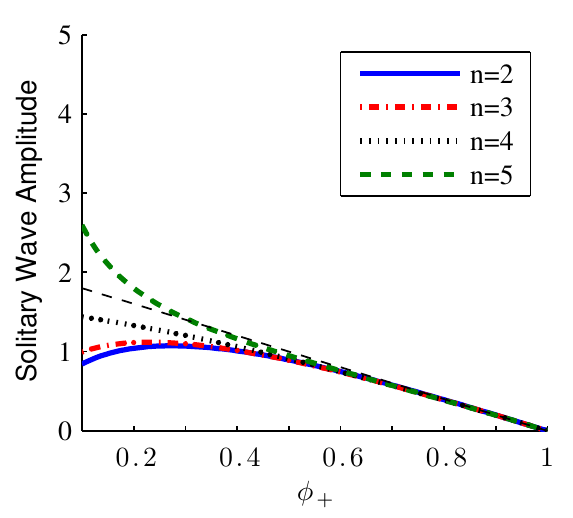}} 
        \subfigure[Solitary Wave Amplitudes, $m=1$]{
        \label{fig: ampsM1}
		\includegraphics{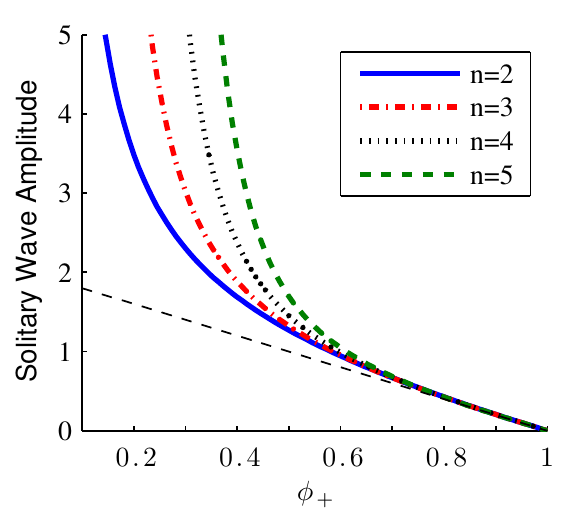}}

        \caption{Analytical predictions for the normalized speeds and amplitude of magma DSWs.  In (a) and (b) the thicker curves indicate the leading edge speeds, and the thinner curves show the trailing edge speeds.  Note the existence of a zero and a universal global minimum in the normalized trailing edge speeds for all cases of $(n,m)$.  For (c) and (d), the thinner dashed line is the KdV amplitude result.}
        \label{fig: analyticalpreds}
\end{figure}

We now consider the speeds for more general parameter values over the full range of $\phi_+$, for which one must solve numerically by implicitly solving the expressions \eqref{eq: implicitalphatrail} and \eqref{eq: betaimplicitlead}.  To solve for the leading edge amplitudes, we invert the amplitude-speed relation \eqref{eq: ampspeed} with a background porosity $\Phi = \phi_+$.   In order to understand the effects of the constitutive parameters $(n,m)$ on DSW behavior, consider figures \ref{fig: speedsM0} and \ref{fig: speedsM1} that show the normalized predicted leading and trailing edge speeds $s_+/n$ and $s_-/n$ of the magma DSW as a function of the downstream porosity $\phi_+$ for $m=0$ and $m=1$, respectively.  The two plots look qualitatively similar indicating that the degree of nonlinearity $n$ appears to have the greatest impact on DSW speeds while $m$ has only a modest effect.  The associated leading solitary wave amplitudes are plotted in figures \ref{fig: ampsM0} and \ref{fig: ampsM1}.  In the amplitude plots, the lighter dashed lines indicate the predictions based on the weakly nonlinear KdV limit.  The amplitudes of the leading edge depend rather dramatically on the choice of the viscosity constitutive parameter $m$.  The inclusion of a porosity weakening matrix viscosity amplifies porosity wave oscillations causing the leading edge to grow large as the jump grows, bounded from below by the KdV amplitude.  In the $m=0$ case, however, the amplitudes grow less rapidly and are approximately bounded above by the KdV amplitude result.  

From the plots of the trailing edge speeds \ref{fig: speedsM0} and
\ref{fig: speedsM1}, we note that not only do the speeds take on
negative values as we discussed in Section \ref{sec: Trailing Edge},
but they also assume a global minimum in the interior of the $\phi_+$
domain. Taking a derivative of eq. \eqref{eq: linspeed} with respect
to $\phi_+$ yields \beqs 
\frac{\ud s_-}{\ud\phi_+} =
\begin{cases}
  \displaystyle
  \frac{2(n-1)n(4\alpha_--1)\phi_+^{2n-3}e^{-2(\alpha_--1)}}{1+2\alpha_-},
  & m+n=2, \\[3mm]
  \displaystyle
  \frac{n(n-m)(4\alpha_--1)\phi_+^{n-m-1}\lp\frac{\alpha_-+M}{1+M}\rp^{1-2M}}
  {1+\alpha_-\lp\frac{2M-1}{\alpha_-+M}\rp}, & \text{otherwise} \prd
\end{cases}	
\eeqs
Then the trailing edge speed derivative is zero only when $\alpha_- =\alpha_\mathrm{min}= \frac{1}{4}$.  Inserting this value of $\alpha$ back into eq. \eqref{eq: linspeed}, we find that the minimum linear speed, for any choice of $(n,m)$ has the universal scaled value 
\beq \label{eq: speed_min}
	\frac{s_-(\alpha_\mathrm{min})}{n} = -\frac{1}{8} \prd
\eeq
We can use the expression \eqref{eq: implicitalphatrail} and $\alpha_-=\frac{1}{4}$ to find that the linear speed takes on a minimum when $\phi_+ = \phi_\mathrm{min}$ where
\beq	\label{eq: phi_min} \phi_\mathrm{min} = 
\begin{cases}
	\displaystyle
	\lp\frac{1}{4}e^{-\frac{3}{2}}\rp^{\frac{1}{2(n-1)}}, & m+n=2 \\[3mm]
	\displaystyle
	  \left\{\frac{1}{4} \left[\frac{1 + 4M}{4(1+M)}\right]^{2M-1}\right\}^{\frac{1}{n-m}}, & 	\text{otherwise}
	\prd	
\end{cases}
\eeq
Note then for $m=1$, we can explicitly verify from eq. \eqref{eq: linspeedM1special} that $\phi_\mathrm{b} > \phi_\mathrm{min}$.  We confirm that this inequality holds for any choice of $(n,m)$ in the valid range.  We will say more about the significance of $\phi_\mathrm{min}$ and its relation with $\phi_\mathrm{b}$ in Section \ref{sec: Breakdown}.


\section{Comparison with Numerical Simulations} \label{sec:
  discussion} The purpose of this section is to compare the analytical
predictions of Section \ref{sec: DSWs} for the speeds and amplitudes
of magma DSWs with careful numerical simulations, as well as to use
simulations to examine the internal shock structure.  We see strong
agreement between predictions and numerics for small to moderate jumps
and identify criteria for the breakdown of the theoretical
construction for large amplitudes.  We find in particular that for
$\phi_+<\phi_\mathrm{min}$, the DSW implodes.  This regime is
characterized by the onset of internal wave interactions corresponding
to gradient catastrophe in the Whitham modulation equations.


\subsection{Numerical Simulations}	\label{sec: numerics}

The magma equation \eqref{eq: magma} with initial data given by \eqref{eq: IC} was simulated using a finite difference spatial discretization and an explicit Runge-Kutta time stepping method.  Details of our numerical method and accuracy verification are found in Appendix \ref{appA}.  Studying numerical simulations allows us not only to verify the analytical predictions described in Section \ref{sec:  DSWs}, but also to study the internal structure of the DSWs, which we sought to bypass before, as well as the limitations of the asymptotic DSW construction

\begin{figure}	
	\centering
        \subfigure[Wine Glass]{
        		\label{fig: wine}
              \includegraphics{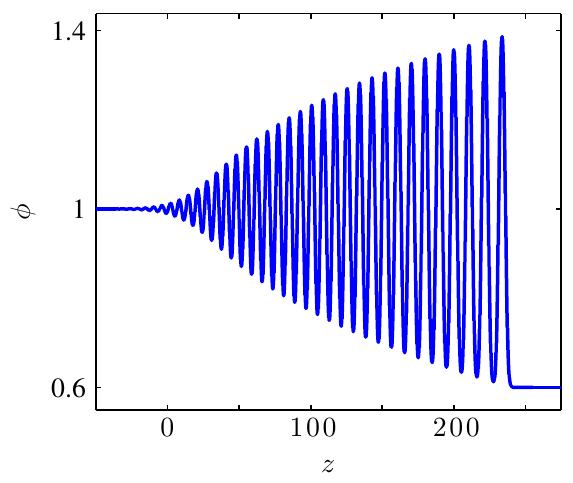}}
	\subfigure[Martini Glass]{
	\label{fig: martini}
                \includegraphics{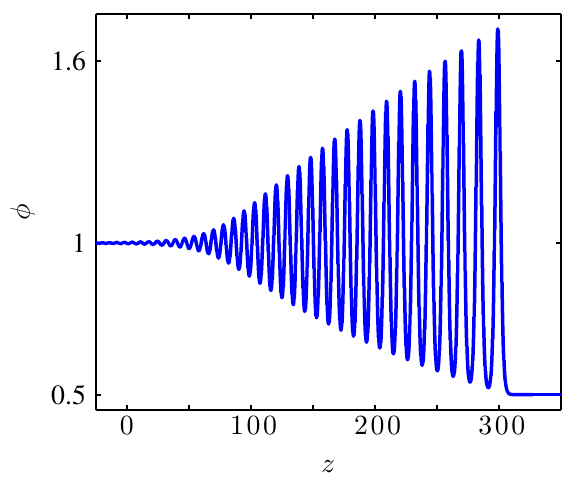}} 
        \caption{Two different types of envelope structures observed in simulations of eq. \eqref{eq: magma}.  (a) The numerical solution at $t=120$ for parameter values $\phi_+=0.6, (n,m)=(3,0)$.  (b) The magma equation at $t=200$ for parameter values $\phi_+ = 0.5, (n,m)=(2,1)$.}
        \label{fig: wine martini}
\end{figure}

We choose to focus on two particular parameter regimes, $(n,m) \in \left\{(2,1), (3,0)\right\}$.  The first case is physically motivated by the fluid conduit problem described in Section \ref{sec: Fluid Problem} while the second was the problem studied numerically by \citet{spiegelman93b} and \citet{marchant05}.  Further, taking different values for $n$ and $m$ in each case allows us to illustrate the different DSW forms arising from eq. \eqref{eq: magma}.  Figure \ref{fig: wine martini} shows the difference in the internal oscillatory structure which results from varying the values of the constitutive parameters.  In language consistent with \citet{kodama_whitham_2008}, figure \ref{fig: wine} when $(n,m)=(3,0)$, depicts a wave envelope in the form of a wine glass, while in figure \ref{fig: martini} with $(n,m)=(2,1)$, the envelope resembles the shape of a martini glass.  The degree of nonlinearity $n$ influences the internal structure of the DSW.  

\begin{figure}
	\centering
        \subfigure[Speeds $(n,m)=(3,0)$]{
        		\label{fig: speedsM0N3}
              \includegraphics{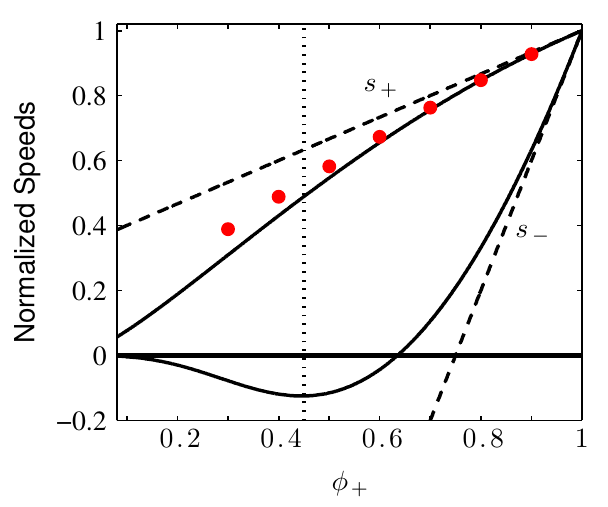}}
	\subfigure[Speeds $(n,m)=(2,1)$]{
	\label{fig: speedsM1N2}
                \includegraphics{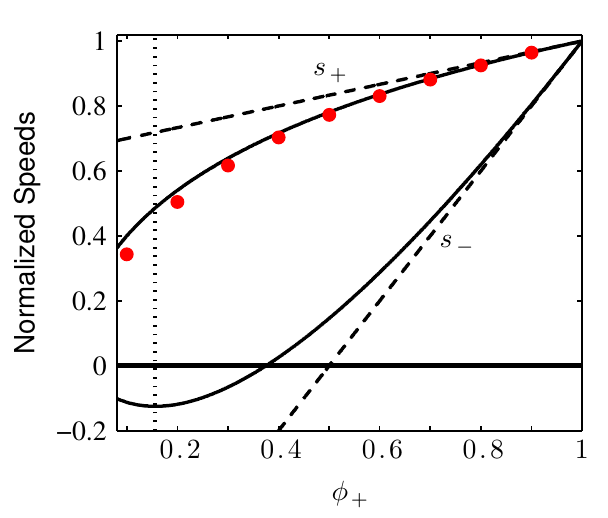}}
        \subfigure[Amplitudes $(n,m)=(3,0)$]{
        \label{fig: ampsM0N3}
                \includegraphics{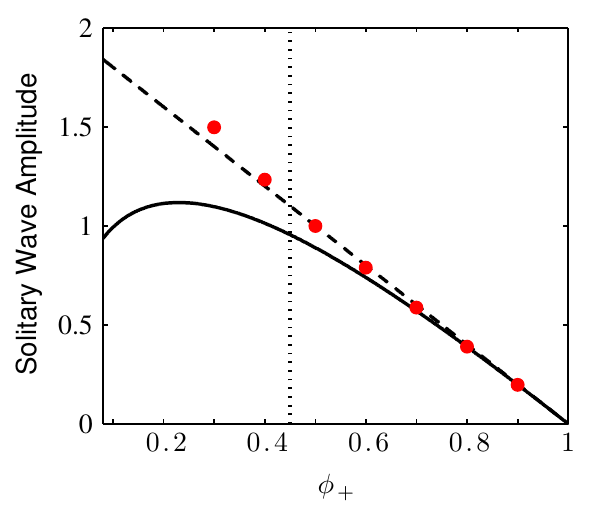}}
        \subfigure[Amplitudes $(n,m)=(2,1)$]{
        \label{fig: ampsM1N2}
		\includegraphics{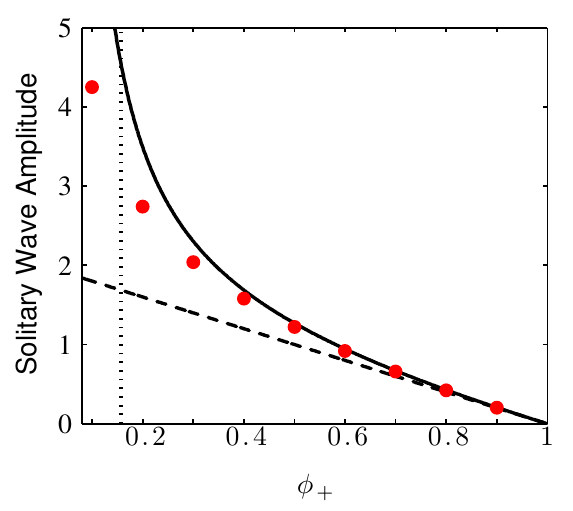}}

      \caption{Comparisons for $(n,m)=(3,0)$ and $(n,m)=(2,1)$ of the
        predicted (solid lines) versus computed values (dot points) of
        the leading edge speeds and amplitudes for varying $\phi_+$.
        The dashed lines indicate the predictions from KdV theory in
        the weakly nonlinear regime.  The vertical dotted line is the
        critical jump height $\phi_\mathrm{c}$, beyond which the
        analytical theory is no longer valid.}
        \label{fig: verify}
\end{figure}

\begin{figure}	
	\centering
        \subfigure[$(n,m)=(3,0), \phi_+ = 0.7$]{
        		\label{fig: contourM0}
              \includegraphics{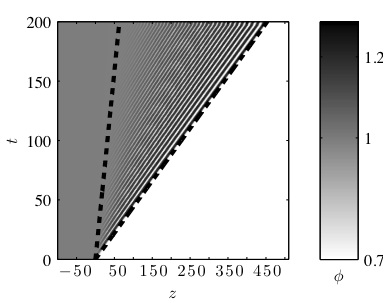}}
	\subfigure[$(n,m)=(2,1), \phi_+ = 0.5$]{
	\label{fig: contourM1}
               \includegraphics{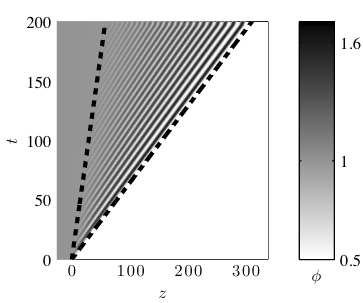}}
             \caption{Contour plots of numerical solutions for
               $\phi(z,t)$ plotted in the characteristic $z$-$t$ plane
               for different parameter values with the overlying
               dashed lines indicating the predicted values for the
               leading and trailing edge speeds.  The angles (speeds)
               line up excellently.}
        \label{fig: contour}
\end{figure}

In figure \ref{fig: verify} we show how the numerically simulated
leading edge speed and amplitude compare with our predicted values.
From figure \ref{fig: speedsM0N3}, we obtain strong agreement for
$\phi_+ \in (0.5,1)$ but increasing deviation thereafter.  Figure
\ref{fig: ampsM0N3} gives a similar picture for the amplitudes, though
curiously the calculated amplitudes appear better predicted by the KdV
asymptotics for larger jump sizes.  From figure \ref{fig: speedsM1N2}
and \ref{fig: ampsM1N2}, we see strong quantitative agreement for
jumps up to $\phi_+ = 0.2$, but then increased disagreement for the
larger jump.  It is interesting to note that in the case
$(n,m)=(3,0)$, the leading edge speed is consistently underestimated,
and thus the leading edge amplitude is as well, while in the case
$(n,m)=(2,1)$ they are consistently overestimated.  We also observe
that good agreement with the speed and amplitude predictions occurs
even for negative trailing edge velocities.  This demonstrates that
backflow is a real physical feature of eq. \eqref{eq: magma}, not just
a mathematical artifact of the solution method.  Note also that while
our numerical simulations for the leading solitary wave may deviate
from the DSW predictions, we verify that the leading edge is indeed a
well-formed solitary wave that satisfies the soliton amplitude-speed
relation.  With the numerically extracted amplitudes we compute the
predicted speed from eq.~\eqref{eq: ampspeed} and compare it with the
speed extracted from the simulations.  The relative errors over all
simulations are less than 0.3~\%.  The trailing edge speeds were more
difficult to extract in an objective, systematic manner due to the
different structures of the trailing edge wave envelope.  Instead, we
show a contour plot of sample solutions in the $z$-$t$ plane for the
two different cases $(n,m)=(3,0)$ in figure \ref{fig: contourM0} and
$(n,m)=(2,1)$ in figure \ref{fig: contourM1}.  Overlying the contour
plots are the predicted DSW region boundary slopes from Section
\ref{sec: DSWs}.  One can see that our predictions are in excellent
agreement with numerical simulations.  We will further validate our
predictions of the linear edge speeds in Section \ref{sec: Breakdown}.

This excellent agreement between predictions and numerical simulations
in the case of small to moderate amplitude jumps has been seen in
other non-integrable physical systems \citep[c.f.][]{esler11, el09}.
However, deviations in the large jump regime are also observed,
leading one to question the validity of the method in the case of
large amplitude.  \citet{el06} posits genuine nonlinearity in the
modulation equations as one possible assumption that could be violated
as the jump size increases.  We will generalize their criteria in the
next section by introducing four verifiable conditions which violate
our analytical construction.

\subsection{Breakdown of Analytical Method}	 	\label{sec: Breakdown}

The Whitham-El DSW simple wave closure method used here requires the
existence of a self-similar simple wave solution to the full set of
the Whitham modulation equations.  In this section, we identify two
mechanisms that lead to the breakdown of this simple wave theory, the
onset of linear degeneracy or zero dispersion.  In the former case, a
loss of monotonicity manifests such that the simple wave can no longer
be continued.  For the latter case, zero dispersion corresponds to a
gradient catastophe in the Whitham modulation equations leading to
compression and implosion of the DSW.

First we consider the zero dispersion case at the trailing edge.  The
right eigenvector of the system \eqref{eq: redsys1}, \eqref{eq:
  redsys2} associated with the characteristic speed
$\frac{\partial}{\partial k}\om_0$ of the trailing edge is
$[0,1]^\mathrm{T}$.  Therefore, in the vicinity of the trailing edge,
the self-similar simple wave satisfies \beqs \bar{\phi}' = 0, \quad k'
= \left ( \frac{\partial^2 \om_0}{\partial k^2} \right )^{-1} \prd
\eeqs Hence, if $\frac{\partial^2}{\partial k^2}\om_0(k,1)\rightarrow
0$ for $k$ near $k_-$, gradient catastrophe in the modulation system
is experienced.  This condition corresponds to a change in the sign of
dispersion and is equivalent to the loss of genuine nonlinearity in
the trailing edge system itself.  As we will show later, this is
distinct from the loss of genuine nonlinearity of the full Whitham
modulation system, in the limit of the trailing edge.  Recalling that
the sign of dispersion changes when $k^2 = 3\Phi^{m-n}$ (see
eq. \eqref{eq: signdisp}), we have the criterion $k_- \ge \sqrt{3}$
leading to gradient catastrophe.  More generally, for any simple wave,
one-phase DSW resulting from step-like initial data of a scalar,
dispersive, nonlinear wave equation for $\phi$, we can formulate the
following condition which must hold for the assumption of a
continuously expanding, one-phase region: \beq \label{eq:
  dispersionsign} \left.\frac{\partial^2}{\partial k^2}\om_0(k,
  \bar{\phi}) \right|_{(k_-, \phi_-)} < 0. \eeq The criterion
\eqref{eq: dispersionsign} holds for systems with negative dispersion
in the long wavelength regime, the inequality reversed for the
positive dispersion case.  This is a natural generalization of the
criterion in modulated \emph{linear} wave theory where a change in the
sign of dispersion is associated with the formation of caustics and
break down of the leading order stationary phase method
\citep{ostrovsky_modulated_2002}.  It is notable that Whitham
hypothesized that breaking of the modulation equations could lead to
an additional source of oscillations (see Section 15.4 in
\citet{whitham74}).  Previous works have resolved breaking in the
Whitham equations by considering modulated multiphase waves in the
context of DSW interactions
\citep{grava_generation_2002,hoefer_interactions_2007,ablowitz_soliton_2009}
or in the context of certain initial value problems \citep{jorge99}.
Beginning with the initial, groundbreaking work \citep{flashka80} on
multiphase Whitham averaging for KdV, all studies since have involved
integrable systems.  Since $k_-$ is a monotonically decreasing
function of $\phi_+$ for the magma DSW, we expect gradient catastrophe
and DSW implosion for sufficiently large jumps, i.e.~$\phi_+ <
\phi_\mathrm{c}$ for the critical jump height $\phi_\mathrm{c}$.
Using our work in Section \ref{sec: DSWs}, we can derive
$\phi_\mathrm{c}$ for which we violate \eqref{eq: dispersionsign}.  We
note the following \beq \label{eq: derivative_phi0}
\frac{\ud}{\ud\phi_+} \left[ \frac{\partial}{\partial k}
  \om_0(k_-,1)\right] = - \left( \frac{\partial^2}{\partial k^2}
  \om_0(k_-,1)\right) k'(1). \eeq The trailing edge wavenumber $k_- =
k(1)$ depends on $\phi_+$ through the initial condition $k(\phi_+) =
0$ (recall eq. \eqref{eq: kint}).  Therefore, in eq. \eqref{eq:
  derivative_phi0}, $\ud k_-/\ud \phi_+ = - k'(1)$.  Hence, the change
in sign of dispersion evaluated at the trailing edge coincides with a
minimum of the linear edge speed as a function of $\phi_+$ (see
figures \ref{fig: speedsM0} and \ref{fig: speedsM1}).  The dispersion
sign changes from negative to positive when $\phi_+=\phi_\mathrm{c} =
\phi_\mathrm{min}$, where $\phi_\mathrm{min}$ is given in
eq. \eqref{eq: phi_min}.  Our analytical method then is no longer
valid for jumps larger than this threshold value.  This coincidence of
a global minimum edge speed and breakdown of the analytical method was
also noted in the case of unsteady, undular bores in shallow-water by
\citet{el06}, however the mechanism was argued to be due to the loss
of genuine nonlinearity in the modulation system.  The sign of
dispersion did not change.

A similar argument holds in the vicinity of the leading edge.  There,
gradient catastrophe occurs if the conjugate phase velocity (i.e. the
speed of the leading edge) increases with decreasing $\tilde{k}$.
Then one-phase behavior is expected to be retained when
\beq \label{eq: solwavespeed} \left.\frac{\partial}{\partial
    \tilde{k}} \left(\frac{\tilde{\om}_\mathrm{s}}{\tilde{k}}\right)
\right|_{(\tilde{k}_+,\phi_+)} > 0 \prd \eeq The condition \eqref{eq:
  dispersionsign} says that the sign of dispersion must remain
negative when evaluated at the trailing edge.  The second condition
\eqref{eq: solwavespeed} requires that the sign of the conjugate
dispersion (now defined through changes in the phase velocity) remain
positive when evaluated at the leading edge. Verifying \eqref{eq:
  solwavespeed}, we find that it does indeed hold for every $\phi_+
\in (0,1)$.  

The simple wave construction also requires that the full modulation
system be strictly hyperbolic and genuinely nonlinear.  Strict
hyperbolicity of the full modulation system requires the three
characteristics be real and distinct at all points except at the DSW
boundaries which correspond to the merger of two characteristics.  In
integrable systems, this can be validated directly due to the
availability of Riemann invariants, but in the non-integrable case we
assume strict hyperbolicity.  Genuine nonlinearity, on the other hand,
is a condition necessary for the construction of the integral curve
connecting the leading and trailing edges and requires that the
characteristic speed $\lambda$ varies \emph{monotonically} along the
integral curve.  Again, we cannot check this condition for all
parameters in the full modulation system, but we can in neighborhoods
near the boundaries.  Parameterizing the integral curve by
$\bar{\phi}$, the monotonicity criteria are \beq \label{eq:
  genNLtrailgen} \left. \frac{\ud \lambda}{\ud\bar{\phi}}
\right|_{(k=k_-, \bar{\phi}=1, a=0)} =
\left. \frac{\partial\lambda}{\partial k} k' + \frac{\partial
    \lambda}{\partial \bar{\phi}} + \frac{\partial \lambda}{\partial
    a} \frac{\ud a}{\ud\bar{\phi}} \right|_{(k=k_-, \bar{\phi}=1,
  a=0)} < 0, \eeq \beq \label{eq: genNLleadgen}
\left. \frac{\ud\lambda}{\ud\bar{\phi}}
\right|_{(\tilde{k}=\tilde{k}_+, \bar{\phi}=\phi_+, \Lambda=0)} =
\left. \frac{\partial \lambda}{\partial \tilde{k}} \tilde{k}' +
  \frac{\partial\lambda}{\partial \bar{\phi}} + \frac{\partial
    \lambda}{\partial \Lambda} \frac{\ud\Lambda}{\ud\bar{\phi}}
\right|_{(\tilde{k}=\tilde{k}_+, \bar{\phi}=\phi_+, \Lambda=0)} > 0
\prd \eeq These monotonicity conditions are the correct way to
determine genuine nonlinearity at the trailing and leading edges.
Testing for genuine nonlinearity in the reduced system of two
modulations equations fails to provide the proper condition (recall
eq. \eqref{eq: dispersionsign}) because they are restricted to the $a
= 0$ plane.  Since the trailing and leading edges correspond to double
characteristics, right and left differentiability, respectively, imply
$\frac{\partial \lambda}{\partial a} \rightarrow 0$ and
$\frac{\partial \lambda}{\partial \Lambda} \rightarrow 0$ at the
appropriate edge.  Also, the limiting characteristic speeds are
$\lambda \rightarrow \frac{\partial}{\partial k}\om_0$ and $\lambda
\rightarrow \frac{\tilde{\om}}{\tilde{k}}$ at the trailing and leading
edges, respectively.  Then the monotonicity criteria \eqref{eq:
  genNLtrailgen} and \eqref{eq: genNLleadgen}, after the change of
variables to $\alpha$ and $\tilde{\alpha}$, simplify to
\begin{equation}
  \label{eq: genNLtrail}
  \begin{split}
    \left.\frac{\ud\lambda}{\ud\bar{\phi}}
    \right|_{(k=k_-, \bar{\phi}=1, a=0)} &=
    \left . \frac{\frac{\partial^2}{\partial k^2} \omega_0
      \frac{\partial}{\partial \bar{\phi}} \omega_0}{n
      \bar{\phi}^{n-1}
    - \frac{\partial}{\partial k} \omega_0} + \frac{\partial^2
    \omega_0}{\partial k \partial \bar{\phi}} \right
|_{(k=k_-, \bar{\phi}=\phi_-, a=0)} \\
    &= -\frac{n\alpha_-\lp
      16\alpha_-^3m-(16m+4)\alpha_-^2+(3n-m+2)\alpha_-+m-1\rp}{2\alpha_-+1}
    < 0,
  \end{split}
\end{equation}
\begin{equation}
  \label{eq: genNLlead}
  \begin{split}
    \left.\frac{\ud\lambda}{\ud\bar{\phi}}
    \right|_{(\tilde{k}=\tilde{k}_+, \bar{\phi}=\phi_+, \Lambda=0)}
    &= \left . \frac{\frac{\partial}{\partial \bar{\phi}} \tilde{\omega}_0 ( \tilde{k}
    n \bar{\phi}^{n-1} - \tilde{\omega}_0)}{\tilde{k}^2 \left (n \bar{\phi}^{n-1} -
      \frac{\partial}{\partial \tilde{k}} \tilde{\omega}_0 \right )}
\right |_{(\tilde{k}=\tilde{k}_+, \bar{\phi}=\phi_+, \Lambda=0)} \\
    &=\frac{\phi_+^{n-2}n\tilde{\alpha} \lp 4\tilde{\alpha}^2 +
      (n-m+4)\tilde{\alpha}+m-1\rp}{2\tilde{\alpha}+1} > 0 \prd
  \end{split}
\end{equation}
When
either of these two conditions is not satisfied, a breakdown in
genuine nonlinearity of the full $3 \times 3$ modulation system occurs
at the boundaries, \eqref{eq: genNLtrail} and \eqref{eq: genNLlead}
corresponding to the trailing, linear edge and the leading solitary
wave edge, respectively.  We can verify analytically that for any
value of $\phi_+ \in (0,1)$ and for all $(n,m)$ in the physical range,
condition \eqref{eq: genNLlead} is satisfied.  For the linear edge
condition \eqref{eq: genNLtrail}, however, we can derive a condition
for $\phi_+$ such that the monotonicity condition is broken.  Linear
degeneracy first occurs when $\alpha = \alpha_\mathrm{l}$, where
$\alpha_\mathrm{l}$ satisfies \beq \label{eq: critfunc}
16m\alpha_\mathrm{l}^3-(16m+4)\alpha_\mathrm{l}^2+(3n-m+2)\alpha_\mathrm{l}+m-1
= 0 \prd \eeq From the IVP \eqref{eq: alphaint} and the implicit
relation between $\alpha$ and $\bar{\phi}$ \eqref{eq:
  implicitalphatrail}, we know that if there exists an
$\alpha_\mathrm{l} \in (0,1]$ which satisfies \eqref{eq: critfunc},
then we can find $\phi_+ = \phi_\mathrm{l}$ such that \eqref{eq:
  genNLtrail} is no longer satisfied.  Then for each $(n,m)$ in the
physical range, there is a critical jump height $\phi_\mathrm{l}$ such
that genuine nonlinearity is lost at the trailing edge.  Note that for
$m=1$, $\alpha_\mathrm{l}=0$ is the only root of eq. \eqref{eq: critfunc} in the
valid range and this gives the value $\phi_\mathrm{l} = 0$ which is
outside the range of interest.  We have verified numerically that
$\phi_\mathrm{c}>\phi_\mathrm{l}$ for all $n\in[2,5]$, $m \in [0, 1]$
so implosion occurs before the loss of genuine nonlinearity.  Before
linear degeneracy occurs, the analytical construction has already
broken down due to a change in sign of dispersion \eqref{eq:
  dispersionsign}.  We have found three distinguished jump heights
exhibiting the ordering
$\phi_\mathrm{b}>\phi_\mathrm{c}>\phi_\mathrm{l}$.  As $\phi_+$ is
decreased through these values, the DSW exhibits backflow and then
implosion before it ever reaches linear degeneracy.

\begin{figure}
	\centering
        \subfigure[$(n,m)=(3,0)$]{
        		\label{fig: breakdownM0N3}
              \includegraphics{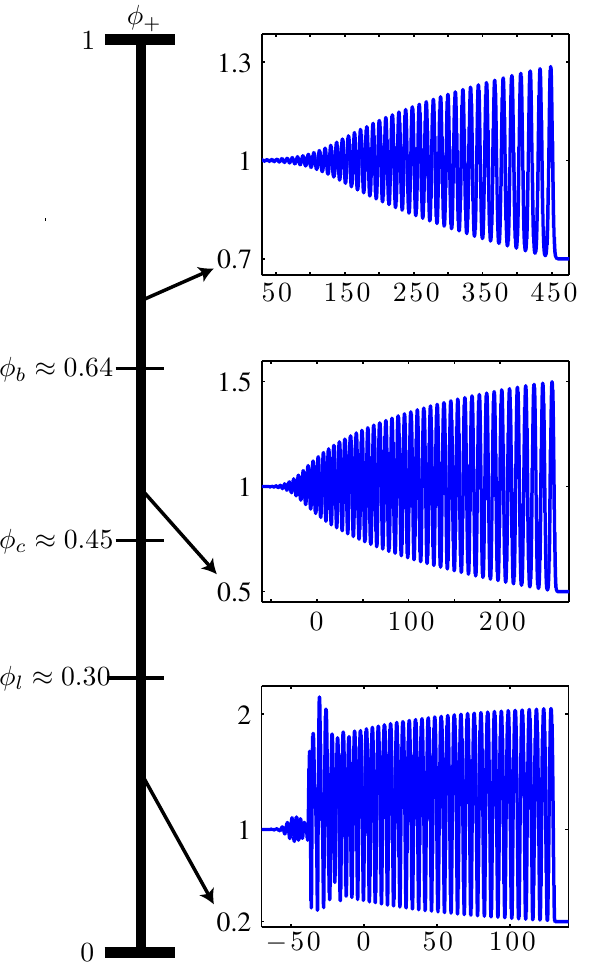}}
	\subfigure[$(n,m)=(2,1)$]{
	\label{fig: breakdownM1N2}
                \includegraphics{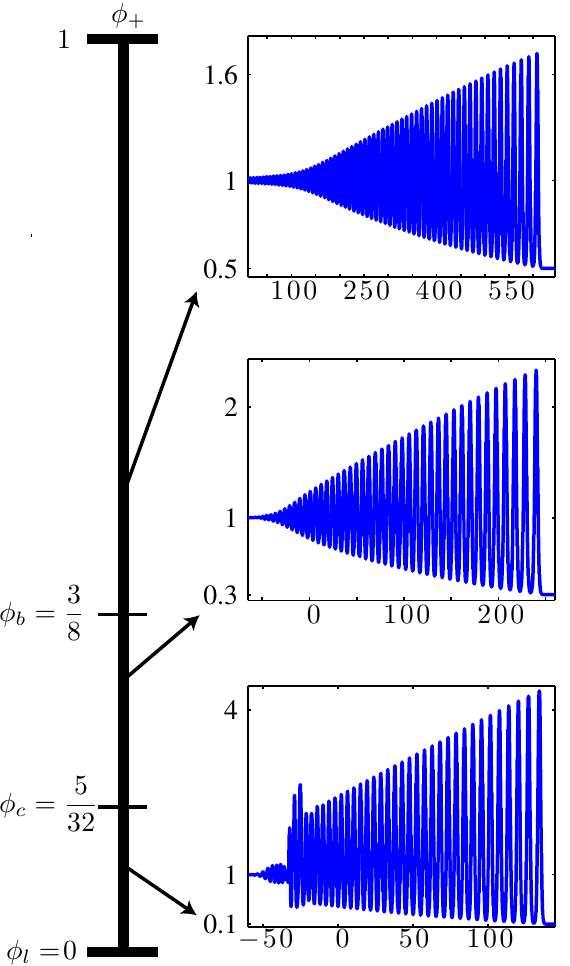}} 
              \caption{From top to bottom, magma DSWs in the forward
                propagating case, the backflow regime before gradient
                catastrophe, and after DSW implosion. The solutions shown correspond, from top
                to bottom, to (a) $t_\mathrm{f} = \left\{200, 150,
                  150\right\}$, (b) $t_\mathrm{f} = \left\{400, 200,
                  200\right\}$.}
        \label{fig: breakdown}
\end{figure}

\begin{figure}
	\centering
    \includegraphics[width=\columnwidth]{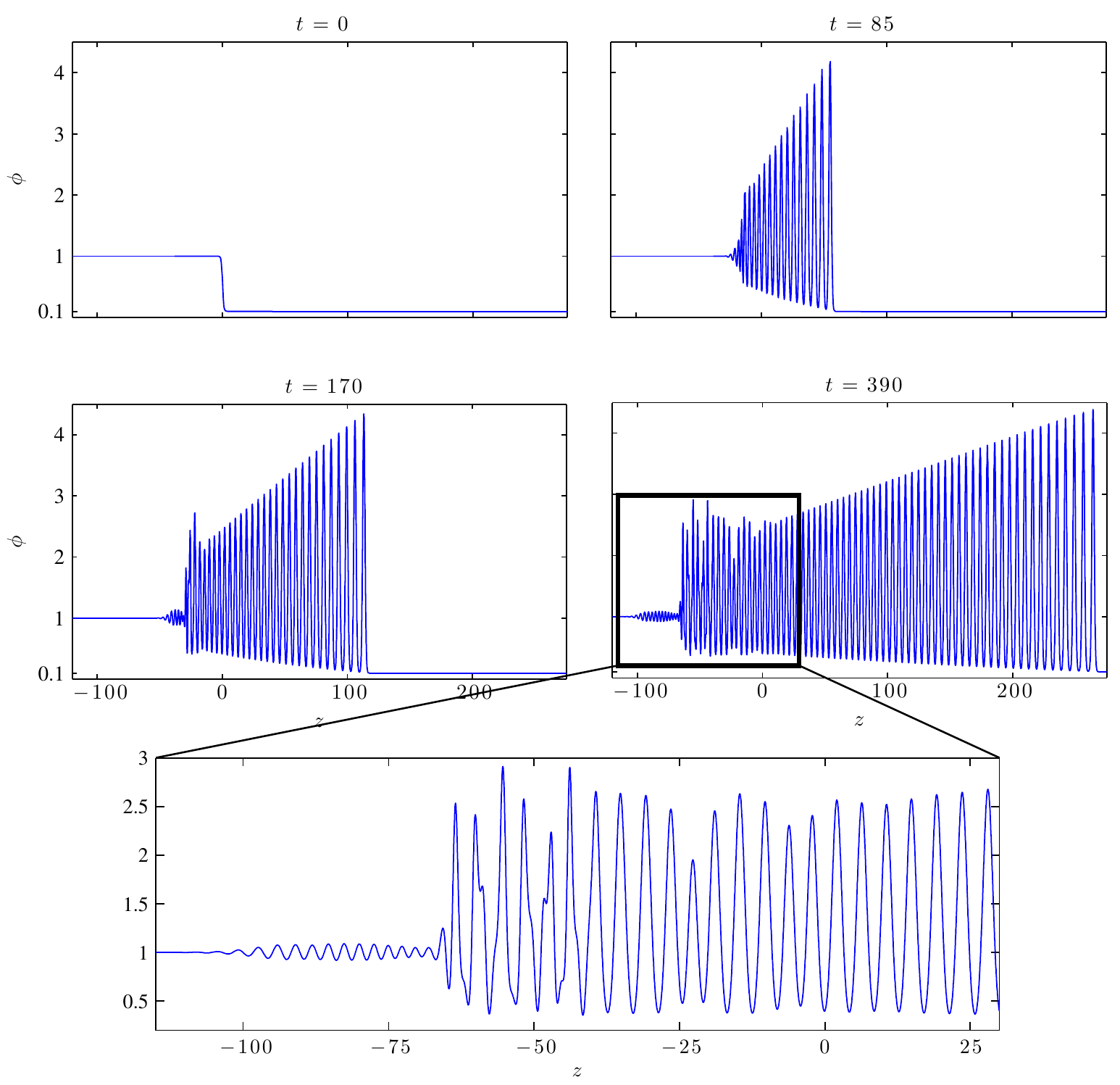}
    \caption{DSW implosion.  Evolution of the solution to
      eq. \eqref{eq: magma} with $(n,m)=(2,1)$, $\phi_+=.1 <
      \phi_\mathrm{c} \approx 0.16$.  The solution initially develops
      into a typical DSW with one-phase interior, but as it evolves,
      the trailing edge compresses and longer waves from the interior
      overtake shorter waves near the edge.  Internal wave
      interactions (implosion) commence and the one-phase assumption
      is no longer valid.}
        \label{fig: catastrophe}
\end{figure}

\begin{figure}
  \centering
  \includegraphics[width=9cm]{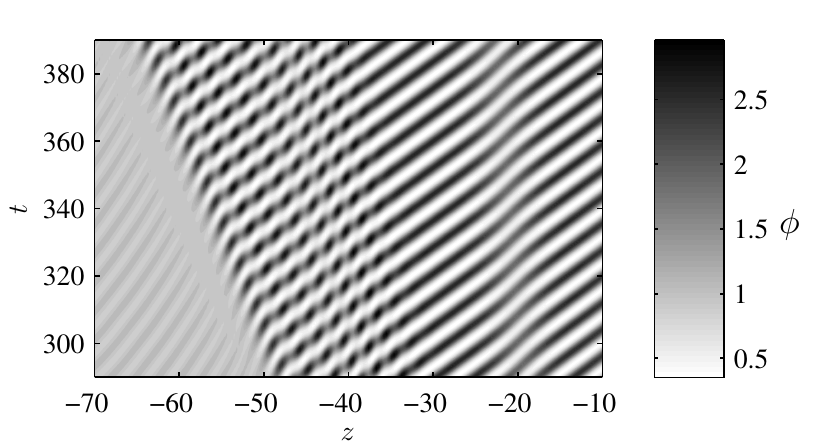}
  \caption{Contour plot of numerically computed $\phi(z,t)$ from
    figure \ref{fig: catastrophe} in the characteristic $z$-$t$ plane
    exhibiting modulated one-phase and two-phase behavior.}
  \label{fig:multiphase}
\end{figure}

Figure \ref{fig: breakdown} shows the key results of our analysis.
Plots of the numerically computed solutions for the two parameter
cases tested in the regimes $\phi_+ \in \left\{ (\phi_\mathrm{b},1),
  (\phi_\mathrm{c},\phi_\mathrm{b}), (0,\phi_\mathrm{c})\right\}$ are
shown.  We see in both cases that our analytical predictions for
$\phi_\mathrm{b}$ and $\phi_\mathrm{c}$ are in agreement with the
numerical solution as backflow and DSW implosion occur when expected
and not otherwise.  The choices for $\phi_+$ in figure \ref{fig:
  breakdown} were chosen for visual clarity but we have performed
further simulations with $\phi_+$ much closer to $\phi_\mathrm{b}$ and
$\phi_\mathrm{c}$, finding that they do indeed accurately predict the
transitions in DSW behavior.  For $\phi_+ < \phi_\mathrm{c}$, the DSW
rank-ordering breaks down due to catastrophe and results in wave
interactions in the interior of the oscillatory region.  In figure
\ref{fig: catastrophe} we show an example of how the solution evolves
from step initial data for $\phi_+<\phi_\mathrm{c}$.  The interior of
the DSW initially develops into approximately a modulated one-phase
region.  However, as the simulation continues, the trailing edge
linear waves stagnate due to the minimum in the group velocity.  The
DSW is compressed as shorter waves at the leftmost edge are overtaken
by longer waves from the interior.  Wave interactions in the DSW
interior ensue.  Another one-phase region develops, separated from the
main trunk of the DSW by a two-phase region, further clarified by the
contour plot of $\phi$ in the characteristic $z$-$t$ plane in figure
\ref{fig:multiphase}.  We find that the closer $\phi_+$ is to
$\phi_\mathrm{c}$, the longer it takes for the interaction region to
develop from step initial data.

This analysis suggests not only that our analytical construction
accurately captures physically important, critical values in $\phi_+$,
but also that our predictions of the trailing edge speed $s_-$ from
Section \ref{sec: Trailing Edge} are consistent with the numerical
simulations.

\section{Summary and Conclusions} \label{sec: summary} In summary, we
have analytically determined DSW speeds and amplitudes for arbitrary
initial jumps and for physical values of the constitutive parameters.
It is likely that these results extend directly to non-physical
parameter values as well.  In figure \ref{fig: breakdown}, we see
backflow and DSW implosion for jumps beyond the catastrophe point
$\phi_\mathrm{c}$, revealing internal wave interactions and the need
for modification of the one-phase region ansatz in the description of
the DSW.  Direct analysis of the full Whitham system would be required
for further study, a highly nontrivial task for arbitrary $(n,m)$.
The significance of this work is its characterization of magma DSW
solutions, an open problem since observed in \citet{spiegelman93b},
its generalization to include the effects of bulk viscosity on
porosity oscillations, and, more generally in the field of dispersive
hydrodynamics, the new DSW dynamics predicted and observed.  A regime
which remains to be understood is the flow of a magma into a ``dry''
region, i.e. $\phi \rightarrow 0$.  The original paper
\citet{spiegelman93b} conjectured that the leading solitary wave would
take on unbounded amplitude, implying physical disaggregation of the
multiphase medium.  This paper has not explored the behavior of
solutions as the jump approaches this singular limit, and a viable
alternative approach is yet to be proposed.

Aside from the novelty of dispersive shock behavior in viscously
dominated fluids, this work provides the theoretical basis with which
to experimentally study DSW generation.  Solitary waves in the fluid
conduit system have been carefully studied experimentally
\citep{scott86b,olson86,helfrich90}, showing good agreement with the
soliton amplitude/speed relation \eqref{eq: ampspeed} in the weak to
moderate jump regime.  DSW speeds, the lead solitary wave amplitue,
and the onset of backflow are all experimentally testable predictions.
Furthermore, the ability to carefully measure DSW properties in this
system would enable the first \emph{quantitative} comparison of
non-integrable Whitham theory with experiment, previous comparisons
being limited to qualitative features
\citep{hoefer_dispersive_2006,wan_dispersive_2007,conti_observation_2009} or
the weakly nonlinear, KdV regime for plasma
\citep{tran_shocklike_1977} and the weakly nonlinear, Benjamin-Ono
equation for atmospheric phenomena \citep{porter02}.  Due to its
experimental relevance, the viscous fluid conduit system deserves
further theoretical study in the large amplitude regime.  What are the
limits of applicability of the magma equation to this system?  How do
higher order corrections affect the dynamics?

Finally, we have established new, testable criteria for the breakdown
of the DSW solution method in equations \eqref{eq: dispersionsign},
\eqref{eq: solwavespeed}, \eqref{eq: genNLtrailgen}, and \eqref{eq:
  genNLleadgen}.  The additional four admissibility criteria--loss of
genuine nonlinearity, change in sign of dispersion at the solitary
wave, linear wave edges--apply to the simple wave DSW construction of
any nonlinear dispersive wave problem.  Linear degeneracy and
nonstrict hyperbolicity have been accommodated in various integrable
nonlinear wave problems
\citep{pierce_self-similar_2007,kodama_whitham_2008,kamchatnov_undular_2012}.
Extensions to non-integrable problems are needed.
\\

We thank Marc Spiegelman for introducing us to the magma equation and
for many helpful suggestions and fruitful discussion.  We appreciate
thoughtful comments from Noel Smyth.  This work was supported by NSF
Grant Nos. DGE--1252376 and DMS--1008973.

\appendix
\section{Numerical method}\label{appA}

The magma equation was simulated using a sixth-order finite difference spatial discretization with explicit Runge-Kutta time stepping.  The initial condition \eqref{eq: IC} was approximated by a smoothed step function centered at $z=0$ 
\beq
	\phi(z,0) = \frac{1}{2} \left[\left(1+\phi_+\right) + \left(\phi_+-1\right)\tanh\left({\frac{z}{4\lambda}}\right)\right],
\eeq
where $\phi$ is the porosity and we assume $\phi(z,t) > 0$ and $\phi_+ \in (0,1)$.  The width $\lambda$ was chosen to be sufficiently large in comparison with the spatial grid spacing, typically $\lambda = 10 \Delta z$ for grid spacing $\Delta z$.  This is reasonable since we are concerned with the long-time asymptotic behavior of the solution and any effects of the initial profile's transients will be negligible.  We choose a truncated spatial domain wide enough in order to avoid wave reflections at the boundary.  It is also convenient to consider the magma equation in the form \eqref{eq: timestepper}, \eqref{eq: elliptic}.  In this form, we have an ODE coupled to a linear elliptic operator $L(\phi) \mathcal{P} = -(\phi^n)_z$.  We first solve for $\mathcal{P}$ by discretizing and inverting $L(\phi)$ using sixth-order, centered differences.  To obtain boundary conditions, we note $\mathcal{P} = \phi^{-m}\phi_t$, and due to the wide domain, the function $\phi$ assumes a constant value at the boundaries so $\mathcal{P}$ decays to zero near the boundaries.  Hence we implement Dirichlet boundary conditions on the compaction pressure.  We then use the solution for $\mathcal{P}$ to update the righthand side of eq. \eqref{eq: timestepper} and step forward in time using the classical, explicit 4th order Runge-Kutta method.  The temporal grid spacing was chosen to satisfy the Courant$-$Friedrichs$-$Lewy (CFL) condition for the dispersionless limit, $\frac{\Delta t}{\Delta z} \le \frac{1}{n}$.  Typically, we took $\Delta t = \frac{\Delta z}{2n}$.

The accuracy of our numerical scheme has been determined by simulating solitary wave solutions on a background porosity $\phi=1$ as derived by \citet{nakayama92}, for the particular parameter regimes $(n,m) \in \{(2,1), (3,0)\}$, used in our analysis.  We numerically solve for $\phi$ from the nonlinear traveling wave equation $(\phi')^2 = g(\phi)$, i.e. eq. \eqref{eq: potential} with $\phi_2 \rightarrow \phi_1$, and use this as our initial profile.  The $\infty$-norm difference between the numerically propagated solution $\phi^*$ and the true solitary wave solution $\phi$ is our figure of merit.  

To compute the initial solitary wave profile, we implicitly evaluate
\beq	\label{eq: solwaveint}
	z-ct = I = \int_\phi^A \frac{\ud\tilde{u}}{\sqrt{g(\tilde{u})}}, \quad z \le 0
\eeq
where $A$ is the peak height of the solitary wave, and $\phi>1$.  We then perform an even reflection about the solitary wave center.  The difficulty in evaluation of \eqref{eq: solwaveint} arises from the square root singularity in the integrand when $\tilde{u}=A$ and the logarithmic singularity when $\tilde{u}$ is near one.  We deal with this by breaking up the integral around these singular points as
\beq
	I = I_1 + I_2 + I_3
\eeq
where 
\beq
	I_1 = \int_\phi^{\phi+\mu} \frac{\ud\tilde{u}}{\sqrt{g(\tilde{u})}}, \quad
	I_2 = \int_{\phi+\mu}^{A-\eps} \frac{\ud\tilde{u}}{\sqrt{g(\tilde{u})}}, \quad
	I_3 = \int_{A-\eps}^A \frac{\ud\tilde{u}}{\sqrt{g(\tilde{u})}},
\eeq
and $\mu>0, \eps>0$ are small and to be chosen.  We evaluate $I_1$ and $I_3$ using Taylor expansions up to first order and calculate $I_2$ via standard numerical quadrature.  The parameters $\mu$ and $\eps$ are chosen so that the approximate error in evaluation of $I_1$, $I_2$, and $I_3$ are less than some desired level of tolerance, $\texttt{tol}$.  Errors in $I_1$ and $I_3$ are due to the local Taylor expansion near the singularities.  The errors in $I_2$ are due to the loss of significant digits in floating point arithmetic.  

For $0<\tilde{u}+\mu-1\ll1$, the approximate rounding error in $I_2$ is
\beq
	fl(\frac{1}{\sqrt{g(\tilde{u}+\mu)}}) 
		\approx (\frac{1}{2}g''(1))^{-\frac{1}{2}}(\tilde{u}+\mu -1 + \eps_{\mathrm{mach}})^{-1},
\eeq
where $fl(x)$ represents the evaluation of $x$ in floating point arithmetic and $\eps_\mathrm{mach}$ is machine precision ($\approx 2.2 \times 10^{-16}$ in double precision).  To constrain the error so that it is less than $\texttt{tol}$, we require
\beq
	\mu >\left(\frac{\eps_{\text{mach}}}{\texttt{tol}\left(2 g''(1)\right)^{\frac{1}{2}}}\right)^{\frac{1}{2}} - \tilde{u}-1,
\eeq
and
\beq
	\eps > \frac{A \ \eps_{\text{mach}}}{2\left(-g'(A)\right)^{\frac{1}{2}}\texttt{tol}}	\prd
\eeq
To evaluate $I_1$, we utilize a first order Taylor expansion of $g(\tilde{u})$ near $\tilde{u}=1$, which gives
\beq
	I_1  \approx \left(\frac{1}{2}g''\left(1\right)\right)^{-\frac{1}{2}} \left[\ln{\left(\frac{\tilde{u}+\mu-1}{\tilde{u}-1}\right)} - \frac{\mu g'''(1)}{6g''(1)}\right]	\prd
\eeq
Then if we take the second term to be the approximate error and require it be less in magnitude than $\texttt{tol}$, we find a second restriction on $\mu$
\beq
	\mu < \frac{6\sqrt{2} \abs{g''(1)}^{\frac{3}{2}} \texttt{tol}}{\abs{g'''(1)}}	\prd
\eeq

A similar procedure for $I_3$ leads to
\beq
	\eps < \left(\frac{6 (-g'(A)) \texttt{tol}}{g''(A)}\right)^{\frac{2}{3}}	\prd
\eeq
Hence, we choose $\eps$ and $\mu$ such that
\beq
	\eps =  \frac{A \ \eps_{\text{mach}}}{2\left(-g'(A)\right)^{\frac{1}{2}}\texttt{tol}},
\eeq
and
\beq
	\mu = \max\left\{0, \left(\frac{\eps_{\text{mach}}}{\texttt{tol}\left(2 g''(1)\right)^{\frac{1}{2}}}\right)^{\frac{1}{2}} - u-1 \right\}	\prd
\eeq

\begin{figure}	
	\centering
        \subfigure[$(n,m)=(3,0)$]{
        		\label{fig: errorM0N3}
              \includegraphics{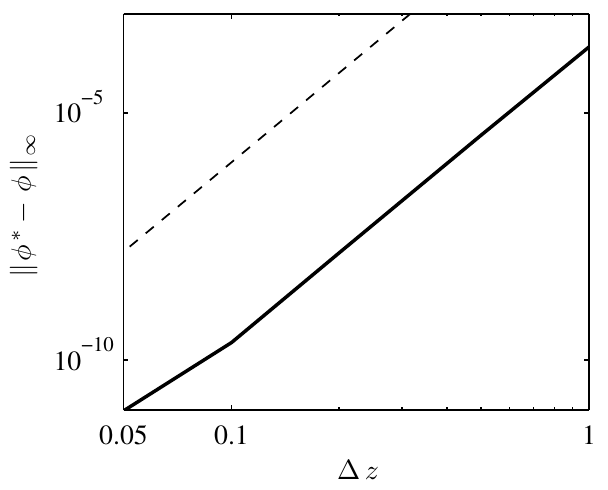}}
	\subfigure[$(n,m)=(2,1)$]{
	\label{fig: errorM1N2}
              \includegraphics{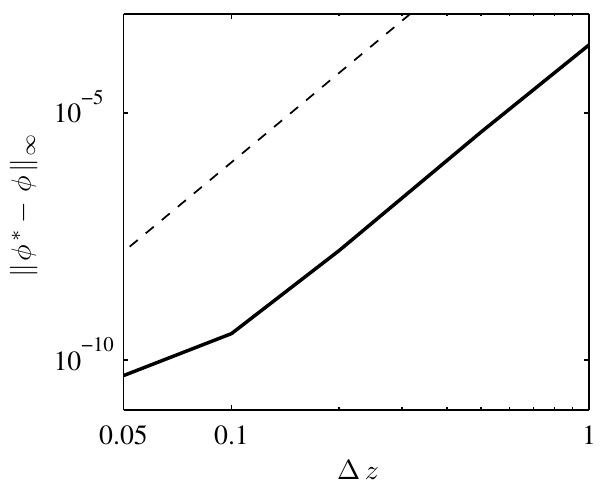}} 
        \caption{Convergence plots of the $\infty$-norm of the difference of the numerically evolved solitary wave versus the true solution.  In both cases, the time discretization $\Delta t$ is fixed and we vary $\Delta z$.  The dotted lines indicate the desired sixth order convergence. }
        \label{fig: error}
\end{figure}

We chose $\texttt{tol} = 10^{-7}$ and found this to give a sufficiently accurate solitary wave profile.  Using this as an initial condition, we initiate our magma equation solver for a solitary wave of height twice the background and a time evolution of approximately 10 dimensionless units.  We ran successive simulations for a range of decreasing $\Delta z$ values and $\Delta t$ chosen as described above (note that simulations of fixed $\Delta z$ and a variable $\Delta t$ showed that the spatial grid was the dominant source of numerical error).  The convergence of the numerical error is described by figure \ref{fig: error}, where the solution converges at sixth order as expected.  Note that there is an alternative method for numerically calculating solitary wave solutions to eq.~\eqref{eq: magma} for arbitrary $n,m$ in all dimensions \citep{simpson11}.  

Based on these solitary wave validation studies, we typically use the
conservative value $\Delta z = 0.05$, though coarser grids were taken
in small amplitude cases where the solutions exhibited longer
wavelength oscillations and had to propagate for much longer times.
The final time $t_\mathrm{f}$ depended upon the jump $\phi_+$, with
$t_\mathrm{f}$ typically at least 1000 for smaller jumps and
$t_\mathrm{f} \approx 200$ for larger jumps.

To calculate the leading edge speed $s_+$ and its amplitude $a_+$, we
generally consider the long time numerical simulations for the last two
dimensionless units of time.  We then find the maximum at each fixed
time and locally interpolate the discrete porosity function on a grid
with spacing $10^{-4}\Delta z$.  This ensures we find the ``true''
numerical maximum and not just the highest point on the grid.  We then
recompute the maximum of the interpolated porosities, and the value at
the final time is the amplitude $a_+$.  To find $s_+$, we compute the
slope of the least squares linear fit to the function connecting the
positions of the interpolated porosity maxima versus their respective
times.  This is the leading edge speed.


\bibliographystyle{jfm}

\begin{thebibliography}{50}
\expandafter\ifx\csname natexlab\endcsname\relax\def\natexlab#1{#1}\fi

\bibitem[Ablowitz {\em et~al.\/}(2009)Ablowitz, Baldwin \&
  Hoefer]{ablowitz_soliton_2009}
{\sc Ablowitz, M.~J., Baldwin, D.~E. \& Hoefer, M.~A.} 2009 Soliton generation
  and multiple phases in dispersive shock and rarefaction wave interaction.
  {\em Phys. Rev. E\/} {\bf 80}, 016603.

\bibitem[Barcilon \& Lovera(1989)]{barcilon89}
{\sc Barcilon, V. \& Lovera, O.~M.} 1989 Solitary waves in magma dynamics. {\em
  J. Fluid Mech.\/} {\bf 204}, 121--133.

\bibitem[Barcilon \& Richter(1986)]{barcilon86}
{\sc Barcilon, V. \& Richter, F.~M.} 1986 Nonlinear waves in compacting media.
  {\em J. Fluid Mech.\/} {\bf 164}, 429--448.

\bibitem[Chanson(2010)]{chanson_tidal_2010}
{\sc Chanson, H.} 2010 {\em Tidal bores, aegir and pororoca: The geophysical
  wonders\/}. World Scientific.

\bibitem[Conti {\em et~al.\/}(2009)Conti, Fratalocchi, Peccianti, Ruocco \&
  Trillo]{conti_observation_2009}
{\sc Conti, C., Fratalocchi, A., Peccianti, M., Ruocco, G. \& Trillo, S.} 2009
  Observation of a gradient catastrophe generating solitons. {\em Phys. Rev.
  Lett.\/} {\bf 102}, 083902.

\bibitem[Dutton {\em et~al.\/}(2001)Dutton, Budde, Slowe \&
  Hau]{dutton_observation_2001}
{\sc Dutton, Z., Budde, M., Slowe, C. \& Hau, L.~V.} 2001 Observation of
  quantum shock waves created with ultra-compressed slow light pulses in a
  {B}ose-{E}instein condensate. {\em Science\/} {\bf 293}, 663.

\bibitem[El(2005)]{el05}
{\sc El, G.~A.} 2005 Resolution of a shock in hyperbolic systems modified by
  weak dispersion. {\em Chaos\/} {\bf 15}, 037103.

\bibitem[El {\em et~al.\/}(2006)El, Grimshaw \& Smyth]{el06}
{\sc El, G.~A., Grimshaw, R.~H.~J. \& Smyth, N.~F.} 2006 Unsteady undular bores
  in fully nonlinear shallow-water theory. {\em Phys. Fluids\/} {\bf 18},
  027104.

\bibitem[El {\em et~al.\/}(2009)El, Grimshaw \& Smyth]{el09}
{\sc El, G.~A., Grimshaw, R.~H.~J. \& Smyth, N.~F.} 2009 Transcritical
  shallow-water flow past topography: finite-amplitude theory. {\em J. Fluid
  Mech.\/} {\bf 640}, 187.

\bibitem[Elperin {\em et~al.\/}(1994)Elperin, Kleeorin \& Krylov]{elperin94}
{\sc Elperin, T., Kleeorin, N. \& Krylov, A.} 1994 Nondissipative shock waves
  in two-phase flows. {\em Physica D\/} {\bf 74}, 372--385.

\bibitem[Esler \& Pearce(2011)]{esler11}
{\sc Esler, J.~G. \& Pearce, J.~D.} 2011 Dispersive dam-break and lock-exchange
  flows in a two-layer fluid. {\em J. Fluid Mech.\/} {\bf 667}, 555--585.

\bibitem[Flashka {\em et~al.\/}(1980)Flashka, Forest \& McLaughlin]{flashka80}
{\sc Flashka, H., Forest, M.~G. \& McLaughlin, D.~W.} 1980 Multiphase averaging
  and the inverse spectral transform of the {K}orteweg--de {V}ries equation.
  {\em Comm. Pure Appl. Math.\/} {\bf 33}, 739--784.

\bibitem[Fowler(1985)]{fowler85}
{\sc Fowler, A.~C.} 1985 A mathematical model of magma transport in the
  asthenosphere. {\em Geophys. Astrophys. Fluid Dynamics\/} pp. 63--96.

\bibitem[Grava \& Tian(2002)]{grava_generation_2002}
{\sc Grava, T. \& Tian, F.-R.} 2002 The generation, propagation, and extinction
  of multiphases in the {KdV} zero-dispersion limit. {\em Comm. Pur. Appl.
  Math.\/} {\bf 55}, 1569--1639.

\bibitem[Gurevich \& Pitaevskii(1974)]{gurevich74}
{\sc Gurevich, A.~V. \& Pitaevskii, L.~P.} 1974 Nonstationary structure of a
  collissionless shock wave. {\em Sov. Phys. JETP\/} {\bf 33}, 291--297.

\bibitem[Harris(1996)]{harris96}
{\sc Harris, S.~E.} 1996 Conservation laws for a nonlinear wave equation. {\em
  Nonlinearity\/} {\bf 9}, 187--208.

\bibitem[Harris \& Clarkson(2006)]{harris06}
{\sc Harris, S.~E. \& Clarkson, P.~A.} 2006 Painleve analysis and similarity
  reductions for the magma equation. {\em SIGMA\/} {\bf 2}, 68.

\bibitem[Helfrich \& Whitehead(1990)]{helfrich90}
{\sc Helfrich, K.~R. \& Whitehead, J.~A.} 1990 Solitary waves on conduits of
  buoyant fluid in a more viscous fluid. {\em Geophys. Astro. Fluid\/} {\bf
  51}, 35--52.

\bibitem[Hoefer \& Ablowitz(2007)]{hoefer_interactions_2007}
{\sc Hoefer, M.~A. \& Ablowitz, M.~J.} 2007 Interactions of dispersive shock
  waves. {\em Physica D\/} {\bf 236}, 44--64.

\bibitem[Hoefer {\em et~al.\/}(2006)Hoefer, Ablowitz, Coddington, Cornell,
  Engels \& Schweikhard]{hoefer_dispersive_2006}
{\sc Hoefer, M.~A., Ablowitz, M.~J., Coddington, I., Cornell, E.~A., Engels, P.
  \& Schweikhard, V.} 2006 Dispersive and classical shock waves in
  {B}ose-{E}instein condensates and gas dynamics. {\em Phys. Rev. A\/} {\bf
  74}, 023623.

\bibitem[Jorge {\em et~al.\/}(1999)Jorge, Minzoni \& Smyth]{jorge99}
{\sc Jorge, M.~C., Minzoni, A.~A. \& Smyth, N.~F.} 1999 Modulation solutions
  for the {B}enjamin--{O}no equation. {\em Physica D\/} {\bf 132}, 1--18.

\bibitem[Kamchatnov {\em et~al.\/}(2012)Kamchatnov, Kuo, Lin, Horng, Gou,
  Clift, El \& Grimshaw]{kamchatnov_undular_2012}
{\sc Kamchatnov, A.~M., Kuo, Y.~H., Lin, T.~C., Horng, T.~L., Gou, S.~C.,
  Clift, R., El, G.~A. \& Grimshaw, R.~H.~J.} 2012 Undular bore theory for the
  {G}ardner equation. {\em Phys. Rev. E\/} {\bf 86}, 036605.

\bibitem[Katz {\em et~al.\/}(2007)Katz, Knepley, Smith, Spiegelman \&
  Coon]{katz07}
{\sc Katz, R.~F., Knepley, M., Smith, B., Spiegelman, M. \& Coon, E.} 2007
  Numerical simulation of geodynamic processes with the {P}ortable {E}xtensible
  {T}oolkit for {S}cientific {C}omputation. {\em Phys. Earth Planet Int.\/}
  {\bf 163}, 52--68.

\bibitem[Kodama {\em et~al.\/}(2008)Kodama, Pierce \&
  Tian]{kodama_whitham_2008}
{\sc Kodama, Y., Pierce, V.~U. \& Tian, F.-R.} 2008 On the {{Whitham}}
  equations for the defocusing complex modified {{KdV}} equation. {\em {SIAM}
  J. Math. Anal.\/} {\bf 41}, 26--58.

\bibitem[Marchant \& Smyth(2005)]{marchant05}
{\sc Marchant, T.~R. \& Smyth, N.~F.} 2005 Approximate solutions for magmon
  propagation from a reservoir. {\em {IMA} J. Appl. Math\/} {\bf 70}, 793--813.

\bibitem[{McKenzie}(1984)]{mckenzie84}
{\sc {McKenzie}, D.} 1984 The generation and compaction of partially molten
  rock. {\em J. Petrol.\/} {\bf 25}, 713--765.

\bibitem[Nakayama \& Mason(1992)]{nakayama92}
{\sc Nakayama, M. \& Mason, D.~P.} 1992 Rarefactive solitary waves in two-phase
  fluid flow of compacting media. {\em Wave Motion\/} {\bf 15}, 357--392.

\bibitem[Olson \& Christensen(1986)]{olson86}
{\sc Olson, P. \& Christensen, U.} 1986 Solitary wave propagation in a fluid
  conduit within a viscous matrix. {\em J. Geophys. Res.\/} {\bf 91},
  6367--6374.

\bibitem[Ostrovsky \& Potapov(2002)]{ostrovsky_modulated_2002}
{\sc Ostrovsky, L.~A. \& Potapov, A.~I.} 2002 {\em Modulated waves: Theory and
  applications\/}. Johns Hopkins University Press.

\bibitem[Pierce \& Tian(2007)]{pierce_self-similar_2007}
{\sc Pierce, V.~U. \& Tian, F.-R.} 2007 Self-similar solutions of the
  non-strictly hyperbolic {Whitham} equations for the {KdV} hierarchy. {\em
  Dyn. {PDE}\/} {\bf 4}, 263--282.

\bibitem[Porter \& Smyth(2002)]{porter02}
{\sc Porter, V.~A. \& Smyth, N.~F.} 2002 Modeling the {M}orning {G}lory of the
  {G}ulf of {C}arpentia. {\em J. Fluid Mech.\/} {\bf 454}, 1--20.

\bibitem[Richter \& {McKenzie}(1984)]{richter84}
{\sc Richter, F.~M. \& {McKenzie}, D.} 1984 Dynamical models for melt
  segregation from a deformable matrix. {\em J. Geol.\/} {\bf 92}, 729--740.

\bibitem[Scheidegger(1974)]{scheidegger74}
{\sc Scheidegger, A.~E.} 1974 {\em The physics of flow through porous media\/}.
  University of Toronto Press.

\bibitem[Scott \& Stevenson(1984)]{scott84}
{\sc Scott, D.~R. \& Stevenson, D.~J.} 1984 Magma solitons. {\em Geophys. Res.
  Lett.\/} {\bf 11}, 1161--1164.

\bibitem[Scott \& Stevenson(1986)]{scott86}
{\sc Scott, D.~R. \& Stevenson, D.~J.} 1986 Magma ascent by porous flow. {\em
  Geophys. Res. Lett.\/} {\bf 91}, 9283--9296.

\bibitem[Scott {\em et~al.\/}(1986)Scott, Stevenson \& Whitehead]{scott86b}
{\sc Scott, D.~R., Stevenson, D.~J. \& Whitehead, J.~A.} 1986 Observations of
  solitary waves in a viscously deformable pipe. {\em Nature\/} {\bf 319},
  759--761.

\bibitem[Simpson \& Spiegelman(2011)]{simpson11}
{\sc Simpson, G. \& Spiegelman, M.} 2011 Solitary wave benchmarks in magma
  dynamics. {\em J. Sci. Comput.\/} {\bf 49}, 268--290.

\bibitem[Simpson {\em et~al.\/}(2007)Simpson, Spiegelman \&
  Weinstein]{simpson07}
{\sc Simpson, G., Spiegelman, M. \& Weinstein, M.~I.} 2007 Degenerate
  dispersive equations arising in the study of magma dynamics. {\em
  Nonlinearity\/} {\bf 20}.

\bibitem[Simpson {\em et~al.\/}(2010)Simpson, Spiegelman \&
  Weinstein]{simpson10}
{\sc Simpson, G., Spiegelman, M. \& Weinstein, M.~I.} 2010 A multiscale model
  of partial melts {I}: Effective equations. {\em J. Geophys. Res.--Sol. Ea.\/}
  {\bf 115}, B04410.

\bibitem[Simpson \& Weinstein(2008)]{simpson_asymptotic_2008}
{\sc Simpson, G \& Weinstein, M.~I.} 2008 Asymptotic stability of ascending
  solitary magma waves. {\em SIAM J. Math. Anal.\/} {\bf 40}, 1337--1391.

\bibitem[Spiegelman(1993{\natexlab{{\em a\/}}})]{spiegelman93a}
{\sc Spiegelman, M} 1993{\natexlab{{\em a\/}}} Flow in deformable porous media
  {I}. simple analysis. {\em J. Fluid Mech.\/} {\bf 247}, 17--38.

\bibitem[Spiegelman(1993{\natexlab{{\em b\/}}})]{spiegelman93b}
{\sc Spiegelman, M.} 1993{\natexlab{{\em b\/}}} Flow in deformable porous media
  {II.} numerical analysis--the relationship between shock waves and solitary
  waves. {\em J. Fluid Mech.\/} {\bf 247}, 39--63.

\bibitem[Spiegelman {\em et~al.\/}(2001)Spiegelman, Kelemen \&
  Aharonov]{spiegelman01}
{\sc Spiegelman, M., Kelemen, P.~B. \& Aharonov, E.} 2001 Causes and
  consequences of flow organization during melt transport: The reaction
  infiltration instability in compactible media. {\em J. Geophys. Res.\/} {\bf
  106}, 2061--2077.

\bibitem[Takahashi {\em et~al.\/}(1990)Takahashi, Sachs \&
  Satsuma]{takahashi90}
{\sc Takahashi, D., Sachs, J.~R. \& Satsuma, J.} 1990 Properties of the magma
  and modified magma equations. {\em J. Phys. Soc. of Jap.\/} {\bf 59},
  1941--1953.

\bibitem[Taylor {\em et~al.\/}(1970)Taylor, Baker \&
  Ikezi]{taylor_observation_1970}
{\sc Taylor, R.~J., Baker, D.~R. \& Ikezi, H.} 1970 Observation of
  collisionless electrostatic shocks. {\em Phys. Rev. Lett.\/} {\bf 24},
  206--209.

\bibitem[Tran {\em et~al.\/}(1977)Tran, Appert, Hollenstein, Means \&
  Vaclavik]{tran_shocklike_1977}
{\sc Tran, M.~Q., Appert, K., Hollenstein, C., Means, R.~W. \& Vaclavik, J.}
  1977 Shocklike solutions of the {K}orteweg-de {V}ries equation. {\em Plasma
  Physics\/} {\bf 19}, 381.

\bibitem[Wan {\em et~al.\/}(2007)Wan, Jia \& Fleischer]{wan_dispersive_2007}
{\sc Wan, W., Jia, S. \& Fleischer, J.~W.} 2007 Dispersive superfluid-like
  shock waves in nonlinear optics. {\em Nat. Phys.\/} {\bf 3}, 46--51.

\bibitem[Whitehead \& Helfrich(1986)]{whitehead86}
{\sc Whitehead, J.~A. \& Helfrich, K.~R.} 1986 The {Korteweg--deVries} equation
  from laboratory conduit and magma migration equations. {\em Geophys. Res.
  Lett.\/} {\bf 13}, 545--546.

\bibitem[Whitham(1965)]{whitham65}
{\sc Whitham, G.~B.} 1965 Non-linear dispersive waves. {\em Proc. R. Soc.
  Lond., Ser. A\/} {\bf 283}, 238--261.

\bibitem[Whitham(1974)]{whitham74}
{\sc Whitham, G.~B.} 1974 {\em Linear and nonlinear waves\/}. Wiley and Sons.

\end{thebibliography}


\end{document}